\documentclass{aa} 
\usepackage[varg]{txfonts}
\usepackage{graphicx}
\usepackage{txfonts}
\usepackage{natbib}
\bibpunct{(}{)}{;}{a}{}{,} 
\begin{document}

\title{On the evolution of ultra-massive white dwarfs}

\author{Mar\'ia E. Camisassa\inst{1,2},
	Leandro G. Althaus\inst{1,2}, 
	Alejandro H. C\'orsico\inst{1,2},
    Francisco C. De Ger\'onimo \inst{1,2},
        Marcelo M. Miller Bertolami\inst{1,2}, 
         Mar\'ia L. Novarino \inst{1},
         Ren\'e D. Rohrmann\inst{3},
         Felipe C. Wachlin \inst{1,2},
    \and  
        Enrique Garc\'\i a--Berro\inst{4,5}*} 
\institute{Facultad de Ciencias Astron\'omicas y Geof\'{\i}sicas, 
           Universidad Nacional de La Plata, 
           Paseo del Bosque s/n, 1900 
           La Plata, 
           Argentina
           \and
           Instituto de Astrof\'isica de La Plata, UNLP-CONICET, 
           Paseo  del Bosque s/n, 1900 
           La Plata, 
           Argentina
     	    \and
           Instituto de Ciencias Astron\'omicas, de la Tierra y del Espacio 
           (CONICET-UNSJ), Av. Espa\~na Sur 1512, J5402DSP, San Juan, Argentina
           \and 
           Departament de F\'\i sica Aplicada, 
           Universitat Polit\`ecnica de Catalunya, 
           c/Esteve Terrades 5, 
           08860 Castelldefels, 
           Spain
           \and
           Institute for Space Studies of Catalonia,  
           c/Gran Capita 2--4, Edif. Nexus 201, 
           08034 Barcelona,  Spain\\
           * Passed away 23th September 2017}
\date{Received ; accepted }

\abstract{{Ultra-massive white dwarfs 
are powerful tools to study various physical processes
in the Asymptotic Giant Branch (AGB), type Ia supernova explosions
and the theory of
crystallization through white dwarf asteroseismology.}
{Despite the interest in these white dwarfs, there are few evolutionary studies in the literature devoted to them.
      Here, we  present  new ultra-massive white dwarf evolutionary  sequences that constitute an 
      improvement over previous ones. }
{In these new sequences, we take into account for 
      the first time the process of phase separation expected during the 
      crystallization stage of these white dwarfs,  by relying  on the most up-to-date
      phase diagram of dense oxygen/neon mixtures. Realistic chemical profiles resulting from the
         full computation of progenitor evolution during the semidegenerate carbon burning along
         the super-AGB phase are also considered in our sequences.  Outer boundary conditions for our
         evolving models are provided by detailed non-gray white dwarf model atmospheres
         for hydrogen and helium composition. }
         {We assessed the impact of all these improvements 
         on the evolutionary properties of ultra-massive white dwarfs, providing up-dated evolutionary
         sequences for these stars.} 
         {We conclude that crystallization is expected to affect the majority of the massive 
         white dwarfs observed with effective temperatures below $40\,000\, \rm K$. Moreover, the 
         calculation of the phase separation process induced by crystallization is necessary to accurately determine
         the cooling age and the mass-radius relation of massive white dwarfs.} {We also provide
colors in the GAIA photometric bands for our H-rich white dwarf evolutionary sequences on the basis of new models atmospheres.}
Finally, these new white dwarf sequences  provide a new theoretical frame
to perform asteroseismological studies on the recently detected ultra-massive pulsating white dwarfs.}
\titlerunning{Ultra-massive white dwarfs}
\authorrunning{Camisassa et al.}  

\maketitle


\section{Introduction}
\label{introduction}

White dwarf stars are the  most common end-point of stellar evolution. Indeed,
more than 97\% of all stars will eventually become white dwarfs. 
These  old  stellar remnants preserve   information about  the
evolutionary  history  of their  progenitors,  providing  a wealth  of
information about  the physical  evolutionary processes of  stars, the
star  formation  history, and  about  the  characteristics of  various
stellar  populations.  
Furthermore,  their  structure  and  evolutionary
properties are well  understood --- see \cite{2008PASP..120.1043F,
2008ARA&A..46..157W},  and   \cite{2010A&ARv..18..471A}  for  specific
reviews --- to the point that the white dwarf cooling  times are currently considered one
of the best age indicators for a wide variety of Galactic populations,
including  open  and  globular clusters  \citep[see][for  some  
applications]{2009ApJ...693L...6W,                2010Natur.465..194G,
2011ApJ...730...35J,     2013A&A...549A.102B,    2013Natur.500...51H}.

The mass distribution of white dwarfs exhibits a main peak at $M_{ \rm WD} \sim 0.6 M_\sun $, and
a smaller peak at the tail of the distribution around $M_{\rm WD} \sim 0.82 M_\sun $
\citep{2013ApJS..204....5K}. 
The existence of massive white dwarfs ($M_{\rm WD} \gtrsim 0.8 M_\sun $) and
ultra-massive white dwarfs  ($M_{\rm WD} \gtrsim 1.10 M_\sun $) 
has been revealed in several studies
\citep{2010MNRAS.405.2561C,2013MNRAS.430...50C,2013ApJ...771L...2H,
2016MNRAS.455.3413K,2017MNRAS.468..239C}. 
Indeed,  \cite{2015MNRAS.452.1637R}
reports the existence of a
distinctive high-mass excess in the mass function of hydrogen-rich white dwarfs near $1\rm M_\sun$.

An historic interest in the study of ultra-massive white dwarfs is related to our understanding of type Ia Supernova.
In fact, it is thought that type Ia Supernova involve the explosion of  an ultra-massive white dwarf or the merger
of two white dwarfs. Also, massive white dwarfs
can act as gravitational lenses. It has been proposed that massive faint white dwarfs can be responsible 
of "microlensing" events in the Large Magallanic Cloud. 

The formation of an ultra-massive white dwarf is theoretically predicted as 
the end product of the isolated evolution of a  massive intermediate-mass star  ---with a mass larger than
6--9 $M_\sun$, depending on metallicity and the treatment of convective boundaries. Once the helium in the core has been exhausted, 
these stars reach the Super Asymptotic Giant
Branch (SAGB) with a  partially degenerate carbon(C)-oxygen(O) core as their less massive siblings.  
However, in the case of SAGB stars  their cores develop temperatures high enough to start carbon ignition under partially degenerate 
conditions. The violent carbon-ignition leads to the formation of a Oxygen-Neon core, which is not hot enough to burn Oxygen or Neon 
(Ne) \citep{2006A&A...448..717S} and is supported by the degenerate pressure of the electron gas. If the  hydrogen-rich 
envelope is removed by winds before electron captures begin in the O-Ne core, an electron-capture supernova is avoided and the 
star leaves the SAGB to form a white dwarf.
As a result, ultra-massive white dwarfs are born with cores composed mainly of $^{16}$O and $^{20}$Ne, with
traces of { $^{12}$C},  $^{23}$Na and $^{24}$Mg \citep{2007A&A...476..893S}. 
In addition, massive white dwarfs with C-O cores can be formed through binary evolution channels; 
namely the single-degenerate channel
 in which a white dwarf gains mass from a nondegenerate companion, and double-degenerate
 channel involving the merger of two white dwarfs \citep{2014ARA&A..52..107M}. 
 The study of the predicted surface properties and cooling times of ultra-massive CO- 
 and ONe-core white dwarfs can help to assess the relevance of different channels in the
 formation of these stars. 

During the last years, $g$(gravity)-mode pulsations have been detected in many massive and 
ultra-massive variable white dwarfs with hydrogen-rich atmospheres (DA), also called ZZ Ceti stars 
\citep{2005A&A...432..219K,2010MNRAS.405.2561C, 2013MNRAS.430...50C,2013ApJ...771L...2H,
2017MNRAS.468..239C}. The ultra-massive ZZ Ceti star BPM 37093 \citep{1992ApJ...390L..89K,2005A&A...432..219K}
was the first object of this kind to be analyzed in 
detail. The existence of pulsating ultra-massive white dwarfs opens the possibility of carrying out asteroseismological analyses of 
heavy-weight ZZ Ceti stars, allowing to obtain information
about their origin and internal structure through 
the comparison between the observed periods and the theoretical periods computed for appropriate theoretical
models. In particular, one of the major interests in the study of pulsating
ultra-massive DA white dwarfs lies in the fact that these stars
are expected to have a well developed crystallized core. 
The occurrence of crystallization in the degenerate core
of white dwarfs, resulting
from Coulomb interactions in very dense plasmas, was first suggested by several authors about 60 yr ago, 
see \citet[][]{Kirzhnits1960,Abrikosov1961,1968ApJ...151..227V} for details, 
and the more recent works by \citet[][]{1999ApJ...526..976M,2004ApJ...605L.133M,
2005A&A...429..277C,2005ApJ...622..572B} for discussions. 
However, this theoretical prediction
was not observationally demonstrated until
the recent studies of \cite{2009ApJ...693L...6W} and \cite{2010Natur.465..194G},
who inferred the existence of crystallized white dwarfs
from the study of the white dwarf luminosity 
function of stellar clusters. Since ultra-massive ZZ Ceti stars 
are expected to have a core partially or totally crystallized, these stars
constitute unique objects to detect the { presence of
crystallization}. 
Thus, asteroseismology of ultra-massive DA
white dwarfs is expected to contribute to our understanding 
of the Coulomb interactions in dense plasmas.
The first attempt to infer the existence of crystallization 
in { an} ultra-massive white dwarf star 
from the analysis of its pulsation pattern was carried out by \citet{2004ApJ...605L.133M} in the case of BPM 37093 \citep
{2005A&A...432..219K}, but the results were inconclusive \citep{2005ApJ...622..572B}.

Asteroseismological applications of ultra-massive DA white dwarfs require the development of 
detailed evolutionary models for these stars, taking into account all the physical processes responsible for 
interior abundance changes as evolution proceeds. The 
first attempts to model these stars by considering 
the evolutionary history of progenitor stars were the studies by  \cite{1997MNRAS.289..973G} and \cite{2007A&A...465..249A}. 
These studies, however, adopted several simplifications
which should be assessed. To begin with, they consider a core chemical profile composed mainly of $^{16}$O and $^{20}$Ne, 
implanted to white dwarf models with different stellar masses. 
A main assumption made in  \cite{2007A&A...465..249A} (from here on, 
we refer to as A07) 
is that the same fixed chemical profile during the entire evolution 
is assumed for all of their models. 
Also, phase separation during crystallization is an important missing physical ingredient in these studies. In fact,  
when crystallization occurs,   energy is released in two different ways. 
First, as in any crystallization process, latent heat energy is released. 
And second, a phase separation of the elements occurs  upon  crystallization, releasing
gravitational energy \citep{1997ApJ...485..308I} and
enlengthening the cooling times of white dwarfs. This process
of phase separation has been neglected in  
all the studies of
ultra-massive white dwarfs.
Finally, progress in the treatment of conductive opacities and model 
atmospheres has been made in recent years,
and should be taken into account in new attempts to
improve our knowledge of these stars.

This paper is precisely aimed at upgrading these old white dwarf evolutionary 
models  by taking into account the above-mentioned considerations. We present new
evolutionary sequences for ultra-massive white dwarfs, 
appropriate for  accurate white
dwarf cosmochronology  of old  stellar systems and for precise asteroseismology of these white dwarfs. 
We compute  four  hydrogen-rich and four hydrogen-deficient white dwarf evolutionary sequences.  
The initial chemical profile of
each white dwarf model is consistent 
with predictions of the progenitor evolution with
stellar masses in the range $9.0\leq M_{\rm ZAMS}/ M_\sun \leq 10.5$
calculated in \cite{2010A&A...512A..10S}.
This chemical structure is the result of the full evolutionary calculations
starting at the Zero Age Main Sequence (ZAMS), and
evolved through the core hydrogen burning, 
core helium  burning, the SAGB  phase, including  the  entire thermally-pulsing phase. 
An accurate nuclear network has been used for each evolutionary phase.
Thus, not only a
realistic O-Ne inner profile is considered for each
white dwarf mass, but also realistic chemical profiles and intershell masses built up
during the SAGB are taken into account. In our study,   the energy released during the crystallization process, as well as the ensuing 
core chemical redistribution were considered by following the phase diagram of \cite{2010PhRvE..81c6107M} 
suitable for $^{16}$O and $^{20}$Ne plasmas\footnote{A. Cumming, personal communication.}.
{ We also provide accurate magnitudes and colors for our hydrogen-rich models
in the filters used by the spacial mission GAIA: G, $\rm G_{BP}$ and $\rm G_{RP}$.}

To the  best of our knowledge, this is  the first set of
fully  evolutionary  calculations  of  ultra-massive white  dwarfs  
including realistic initial chemical profiles for each white dwarf mass, 
an updated microphysics, and the effects of phase separation process duration  crystallization\footnote{These evolutionary
sequences are available at {\tt http://evolgroup.fcaglp.unlp.edu.ar/TRACKS/ultramassive.html}}.
This paper  is organized  as follows.   In Sect.~\ref{code}  we briefly
describe  our  numerical  tools  and   the  main  ingredients  of  the
evolutionary  sequences, while  in Sect.~\ref{results}  we present  in
detail our evolutionary results and compare them with previous works. Finally,  in
Sect.~\ref{conclusions} we  summarize the main findings  of the paper,
and we elaborate on our conclusions.

\section{Numerical setup and input physics}
\label{code}

The white dwarf evolutionary   sequences  presented  in  this   work  have  been
calculated using  the {\tt LPCODE}  stellar evolutionary code  
\citep[see][for  details]{2005A&A...435..631A,
2012A&A...537A..33A}. 
This code has been well tested and calibrated and has been amply used in
the  study  of  different  aspects of  low-mass  star  evolution
\citep[see][and references  therein]{2010Natur.465..194G,
2010ApJ...717..897A,2010ApJ...717..183R}.  More  recently,
the code has been  used to generate a new grid  of models for post-AGB
stars  \citep{2016A&A...588A..25M}
and also new evolutionary sequences for hydrogen-deficient white dwarfs \citep{2017ApJ...839...11C}. 
We mention that {\tt  LPCODE}  
has  been tested  against another  white
dwarf evolutionary code, and the uncertainties  in the white dwarf cooling
ages  that result from the  different  numerical  implementations of  the
stellar   evolution   equations   were   found   to   be   below   2\%
\citep{2013A&A...555A..96S}.

For the white dwarf regime, the main input physics of {\tt LPCODE} includes the
following ingredients.  Convection is  treated within  the
standard   mixing   length   formulation,   as  given   by   the   ML2
parameterization \citep{1990ApJS...72..335T}. 
Radiative   and    conductive   opacities   are     from   OPAL
\citep{1996ApJ...464..943I}   and   from   \cite{2007ApJ...661.1094C},
respectively.    For the low-temperature regime, molecular
radiative opacities with varying carbon to oxygen ratios are used.  To this end,
the   low  temperature   opacities  computed   by
\cite{2005ApJ...623..585F}         as         presented          by
\cite{2009A&A...508.1343W} are adopted.   
The  equation  of   state  
for the low-density regime is taken from \cite{1979A&A....72..134M}, whereas
for  the high-density  regime,  we  employ the  equation  of state  of
\cite{1994ApJ...434..641S}, which includes  all  the  important
contributions for both the solid  and liquid phases. 
We considered neutrino emission
for pair,  photo,  and bremsstrahlung  processes using the rates of
\cite{1996ApJS..102..411I}, while  for plasma processes we  follow the
treatment presented in \cite{1994ApJ...425..222H}. { Outer boundary conditions for both H-rich and H-deficient evolving models are provided
by non-gray model atmospheres, see \cite{2012A&A...546A.119R}, \cite{2017ApJ...839...11C}, and \cite{2018MNRAS.473..457R} for references. The impact of the atmosphere treatment on the cooling times becomes relevant for effective temperatures
lower than $10\,000$ K. }
{\tt LPCODE} considers a detailed treatment of element
diffusion, including gravitational settling, chemical and thermal diffusion. As we will see,
element diffusion is a key ingredient in shaping the chemical profile of evolving
ultra-massive white dwarfs, even in layers near the core.

\subsection{Treatment of crystallization}

 A main issue in the modelling of ultra-massive white dwarfs is the treatment of crystallization. As  temperature decreases in the interior of white dwarfs,
the Coulomb interaction energy becomes increasingly important, until at some point,
they widely exceed the thermal motions and the ions begin to freeze into a regular lattice structure. 
Since the crystallization temperature of pure  $^{20}$Ne is larger than the crystallization temperature of $^{16}$O, 
this crystallization process induces a phase separation. In a mixture of $^{20}$Ne and
 $^{16}$O, the crystallized plasma will be enriched in $^{20}$Ne and, consequently, $^{20}$Ne will decrease in the remaining liquid plasma.
{ This process releases gravitational energy, thus constituting a new energy source that will impact the cooling times.}

\begin{figure}
\centering
\includegraphics[clip,width=\columnwidth]{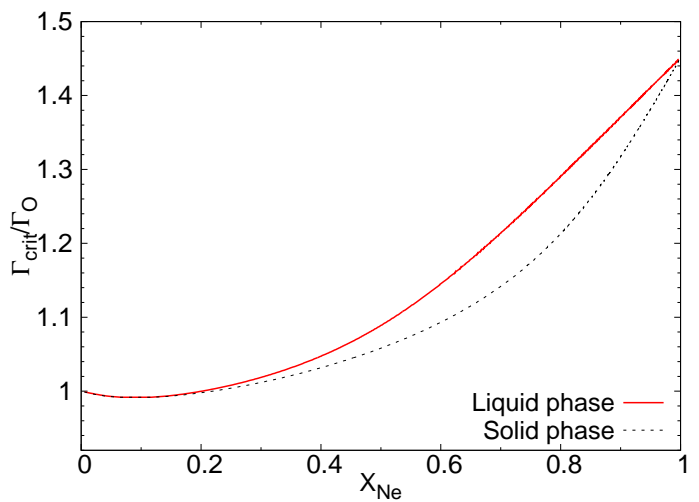}
\caption{Phase diagram of crystallization for a $^{16}$O/$^{20}$Ne mixture \citep{2010PhRvE..81c6107M}. 
$\rm X_{Ne}$ is
the $^{20}$Ne abundance. 
$\rm \Gamma_{crit}$ is set to 178.6. $\rm \Gamma_O$ 
is given by $ \Gamma_{\rm O}=(e^2 / \rm k_B a_e T)8^{5/3}$, see text for details.}
\label{Fig:PD}
\end{figure}

{ We used} the most up-to-date phase diagram   
of dense O-Ne
mixtures  appropriate for massive white dwarf  interiors  \citep{2010PhRvE..81c6107M}. 
This phase diagram, shown in Fig. \ref{Fig:PD}, yields the temperature at
which crystallization occurs, as well as  the abundance change at a given point 
in the solid phase during the phase transition.
$\Gamma$ is the coulomb coupling parameter, defined as  
$\Gamma=\frac{\rm e^2}{{\rm k_B a_e }T}Z^{5/3}$,
where $\rm a_e= \left( \frac{3}{4\pi n_e}\right)^{1/3}$
is the mean electron spacing.
$\rm \Gamma_{crit}$ is set to 178.6,
the crystallization value 
of a mono-component plasma. 
$ \rm \Gamma_O$ 
is the value of $\Gamma$  of $^{16}$O at which crystallization of the mixture occurs, and is related to the temperature and the density
through the relation $ \Gamma_{\rm O}=\frac{\rm e^2}{{\rm k_B a_e }T}8^{5/3}$.
For a given mass fraction of $^{20}$Ne, the solid red line in Fig. \ref{Fig:PD} gives us  $\rm \Gamma_O$,
and, consequently, the temperature of crystallization is obtained.
Once we obtain this temperature, it can be related to the
$\Gamma$ of the mixture, by replacing $T$ in the formula
$\Gamma=\frac{\rm e^2}{{\rm k_B a_e }T}Z_{\rm mixture}^{5/3}$, where 
$Z_{\rm mixture}$ is the mean ionic charge of the mixture.
The $\Gamma$ obtained using this procedure is larger than the value of 
$\Gamma$ commonly used in the white dwarf evolutionary calculations,
which is artificially set to 180. 
For a given abundance of
$^{20}$Ne in the liquid phase, the solid red line predicts $\rm \Gamma_{crit}/\Gamma_{O}$,
and the corresponding value of $\rm \Gamma_{crit}/\Gamma_{O}$ at the dashed black line
predicts the $^{20}$Ne abundance in the solid phase, which is slightly larger than the initial $^{20}$Ne abundance.
The final result of the crystallization process is that the inner regions of the star
are enriched in $^{20}$Ne, and the outer regions are enriched in $^{16}$O. 

\begin{figure*}
\centering
\includegraphics[clip,width=2\columnwidth]{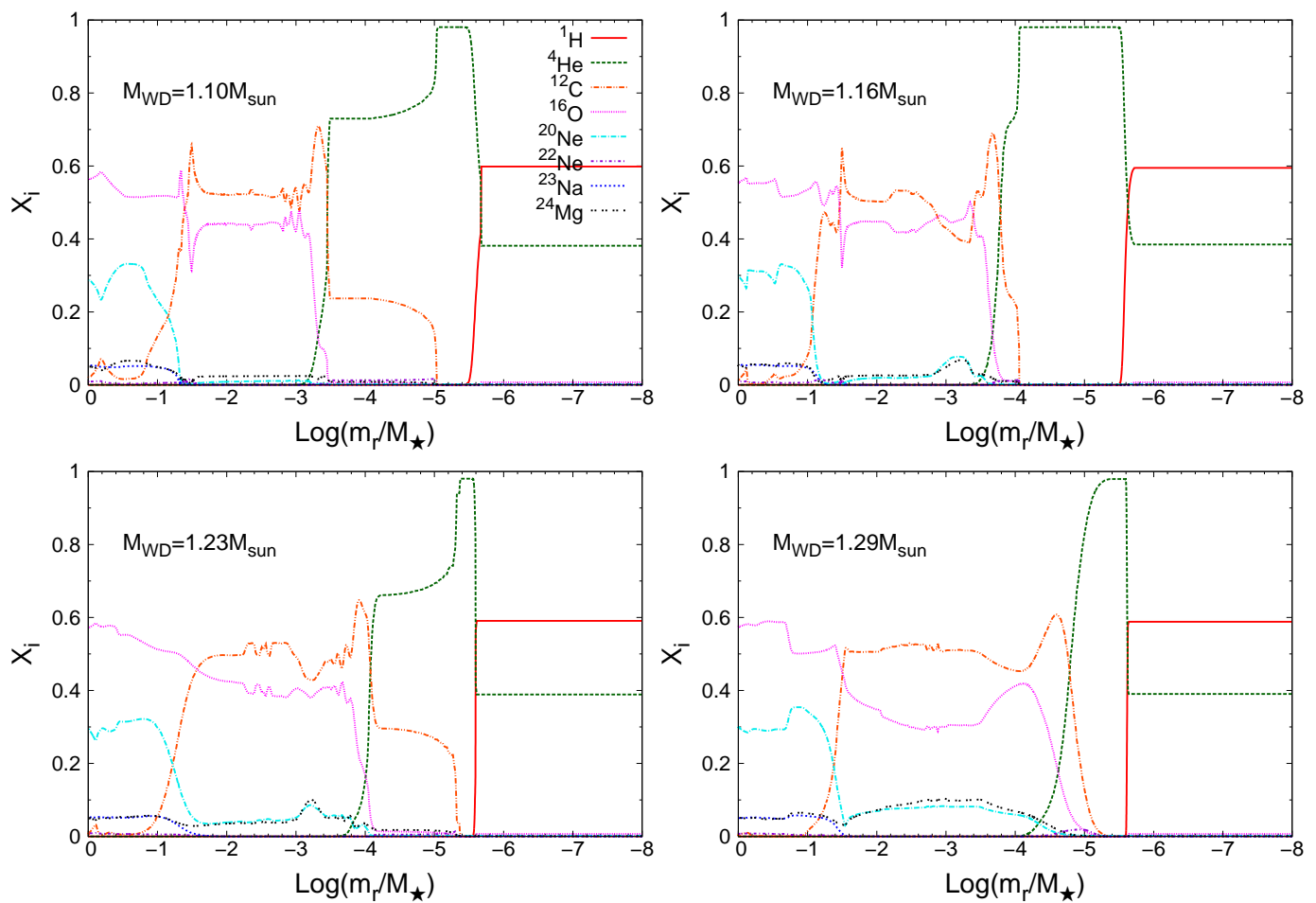}
\caption{Initial chemical profiles before Rayleigh-Taylor 
rehomogeneization corresponding to our four hydrogen-rich white dwarf models as given by nuclear history of progenitor stars.}
\label{Fig:profiles}
\end{figure*}

\begin{figure*}
\centering
\includegraphics[clip,width=2\columnwidth]{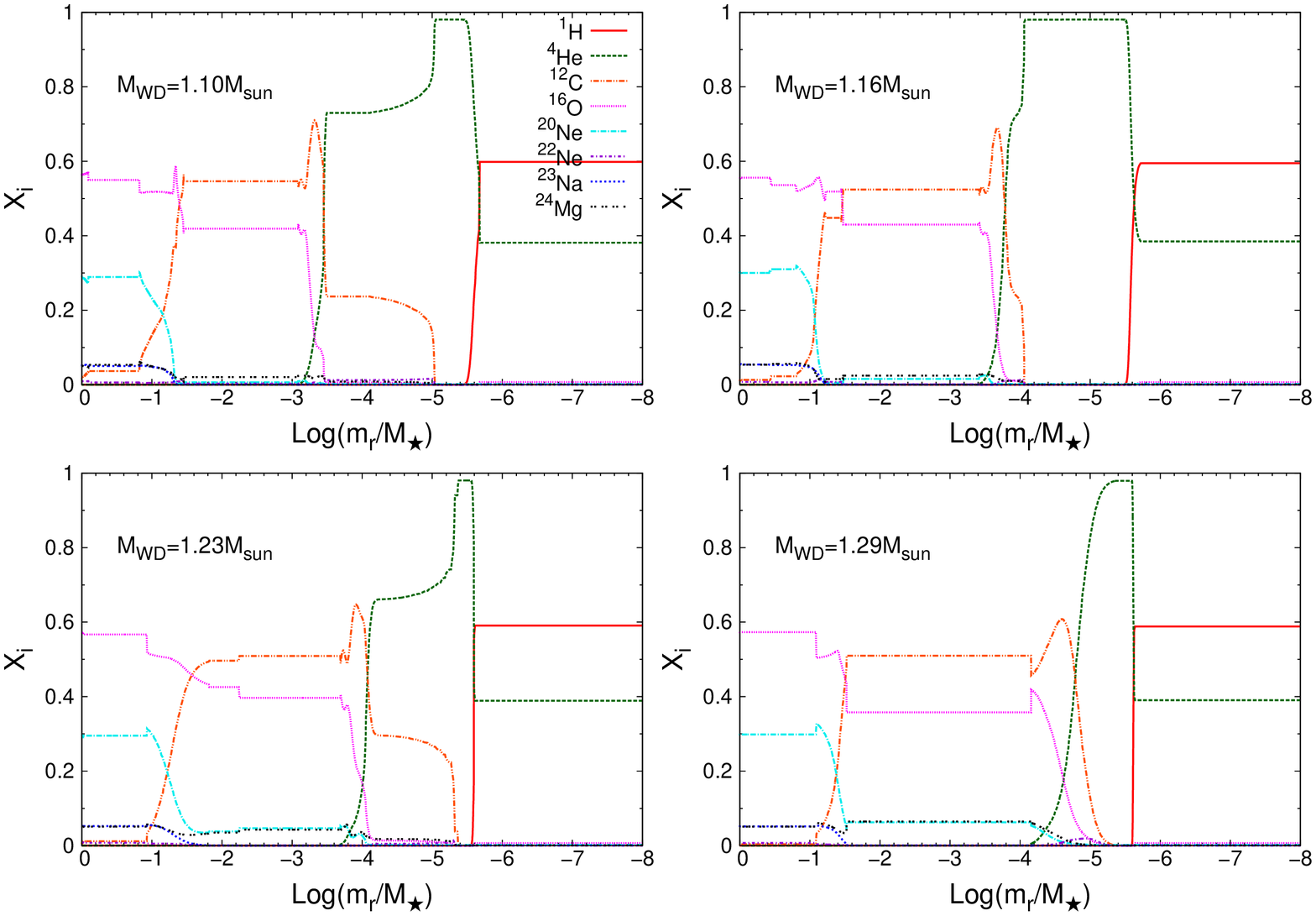}
\caption{Initial chemical profiles of our four hydrogen-rich white dwarf models
once Rayleigh-Taylor rehomogeneization has occurred.}
\label{Fig:profiles2}
\end{figure*}

The energetics resulting from crystallization processes has been 
self-consistently and locally coupled to the full  set of  equations of stellar  evolution
\citep[see][for details of the implementation]{2010ApJ...719..612A}. 
The local change of chemical abundance resulting from the process of phase separation at 
crystallization leads to a release of energy (in addition 
to the latent heat).
The  inclusion  of this energy in {\tt LPCODE} is similar to that described in
\cite{2010ApJ...719..612A}, but
adapted to the mixture of $^{16}$O and $^{20}$Ne characterizing the  core of our ultra-massive
white dwarf models. At each
evolutionary  time  step,  we  calculate  the  change  in chemical
composition  resulting  from   phase  separation
using  the  phase  diagram of \cite{2010PhRvE..81c6107M} for an oxygen-neon
mixture. Then, we evaluate the
net energy released by this process during the time step. This energy is added to the latent
heat contribution, which is considered as $0.77 k_BT$ per ion. The total energy is
distributed  over  a  small  mass  range  around  the
crystallization front. This local energy contribution is added
to the luminosity equation \citep[see][for details]{2010ApJ...719..612A}.

{ The increase of $^{20}$Ne abundance in the solid core as a result of crystallization leads} to a Rayleigh-Taylor instability and an ensuing mixing 
process 
at  the  region  above  the  crystallized  core, inducing 
the oxygen enrichment
in the overlying  liquid  mantle \citep{1997ApJ...485..308I}.
Thus, those  layers  that  are  crystallizing contribute as an 
energy source,  and the  overlying unstable layers will be a  sink  of  energy.


\subsection{Initial models}

As we have mentioned, an improvement of the present calculations over those published 
in A07 is the adoption of detailed chemical profiles which are based on the computation of all the previous evolutionary stages of their 
progenitor stars. This is true for both the O-Ne core and the surrounding envelope. In particular, the full computation of previous
evolutionary stages allows us to assess the mass of the helium-rich mantle and the hydrogen-helium transition, which are of particular 
interest for the asteroseismology of ultra-massive white dwarfs. Specifically, the chemical
composition of our models is the result of the entire progenitor evolution calculated  in \cite{2007A&A...476..893S, 2010A&A...512A..10S}. 
These sequences correspond to the 
complete single evolution from the ZAMS to the thermally pulsating SAGB phase of initially 
$M_{\rm ZAMS}= 9, 9.5, 10$, and $10.5 M_\sun$
sequences with an initial metallicity of $Z= 0.02$. Particular care was taken by
\cite{2007A&A...476..893S, 2010A&A...512A..10S} to precisely
follow the propagation of the carbon burning flame where most carbon is burnt \citep{2006A&A...448..717S}.
This is of special interest for the 
final oxygen and neon abundances in the white dwarf core. In addition, \cite{2010A&A...512A..10S}
computed in detail the evolution during the
thermally pulsing-SAGB phase where the outer chemical profiles and the total helium-content of the
final stellar remnant are determined.
{ No} extra mixing was included at any convective boundary at any evolutionary stage. The absence
of core overshooting 
during core hydrogen- and helium-burning stages implies that, for a given final remnant mass ($M_{\rm WD}$), 
initial masses ($M_{\rm ZAMS}$) 
represent an upper limit of the expected progenitor masses.
{ Indeed, considering moderate overshooting during core helium burning 
lowers
the mass range
of SAGB stars in $2\, \rm M_\odot$ \citep{2007A&A...476..893S,2007A&A...464..667G}.}
It is worth noting that the initial 
final mass relation is poorly constrained from 
observations \citep{2009ApJ...692.1013S} and it is highly uncertain in stellar evolution models.
{ On the other hand, considering overshooting during the thermally-pulsing SAGB, would induce
third dredge-up episodes, altering the carbon and nitrogen abundances in the envelope. Finally, in this work we have not explored the impact on white dwarf cooling that could be expected from changes in the core chemical structure resulting from the consideration of extra- mixing episodes during the semi-degenerate carbon burning.}

The stellar masses of our white dwarf sequences are $M_{\rm WD}=1.10 M_\sun$, $1.16  M_\sun$, 
$1.23 M_\sun$ and $1.29  M_\sun$. Each evolutionary sequence was computed from the beginning of the cooling track at high luminosities down to the development of
the full Debye cooling at very low surface
luminosities, $\log(L_\star/L_\sun)= -5.5$.
 The progenitor evolution through the thermally-pulsing SAGB provides us with  realistic values of the
total helium content, which is relevant for accurate computation 
of cooling times at low luminosities. In particular, different helium masses lead to different cooling times.
The helium mass of our $1.10  M_\sun$, $1.16 M_\sun$, 
$1.23M_\sun$ 
and $1.29  M_\sun$ models are $3.24 \times 10^{-4} M_\sun$,
 $1.82 \times 10^{-4} M_\sun$,  $0.78 \times 10^{-4}  M_\sun$ and
  $0.21 \times 10^{-4} M_\sun$, respectively. By contrast, the total mass of the 
  hydrogen envelope left by prior evolution is quite uncertain, since it depends on the
  occurrence of carbon enrichment on the
  thermally pulsing AGB phase \citep[see][]
  {2015A&A...576A...9A}, which in turn 
  depends on the amount of overshooting and mass loss, as well as on the occurrence of
  late thermal pulses. For this paper, we have adopted the maximum expected hydrogen envelope 
  of about  $\sim 10^{-6} M_{\sun}$ for  ultra-massive white dwarfs. Larger values of the total
  hydrogen mass would lead to unstable nuclear burning and thermonuclear flashes on the white dwarf
  cooling track.

Fig. \ref{Fig:profiles} illustrates the chemical profiles resulting from the progenitor evolution 
of our four hydrogen-rich white dwarf sequences\footnote{The chemical profiles of our hydrogen-deficient white dwarf models are the same, 
except that no hydrogen is present in the envelope.}.
The core composition is $\sim 55\%$ $^{16}$O, $\sim 30\%$ $^{20}$Ne, with 
minor traces of $^{22}$Ne,
$^{23}$Na, $^{24}$Mg.
{  At some layers of the models, the mean molecular weight is higher than in the deeper layers, leading to Rayleigh-Taylor unstabilities. Consequently,
these profiles are expected to 
 undergo a rehomogeneization process in a timescale shorter than the evolutionary timescale. Thus,
we have simulated the rehomogeneization process assuming to be instantaneous. The impact of this mixing process on the
abundance distribution in the white dwarf core results apparent
from inspecting Fig. \ref{Fig:profiles2}. Clearly, rehomogeneization  
mixes the abundances of all elements at some layers of core, erasing preexisting peaks in the abundances.}

\section{Evolutionary results}
\label{results}

\begin{figure}
\centering
\includegraphics[clip,width=\columnwidth]{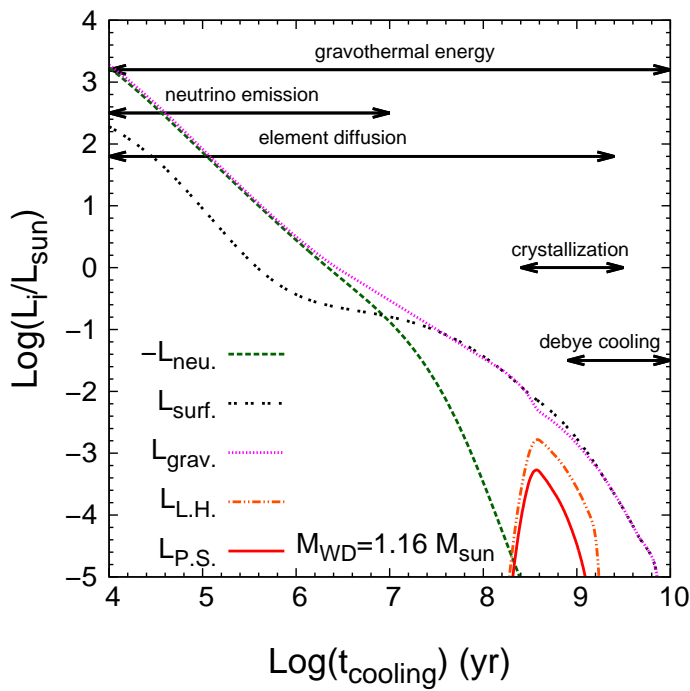}
\caption{ Temporal evolution of surface luminosity (double dotted line) and different
luminosity contributions: neutrino luminosity (dashed line),
gravothermal luminosity (dotted line),  latent heat (dot-dot-dashed line) and phase separation energy (solid line). The arrows indicate the main physical processes responsible for the evolution at different moments.}
\label{Fig:4}
\end{figure}

 We { present} in Fig.~\ref{Fig:4} a global view  of  the 
 main  phases  of the  evolution  of 
an ultra-massive hydrogen-rich white  dwarf model during the  cooling phase. 
In this figure, the temporal evolution of the different luminosity contributions
is displayed for our $1.16 M_{\sun}$ hydrogen-rich white dwarf sequence. 
The cooling time is defined  as zero at the
beginning of the white dwarf cooling  phase, when the star reaches the
maximum effective temperature.
During  the entire white
dwarf evolution,  the release of  gravothermal energy is  the dominant
energy source of the star.  
At early stages, neutrino emission constitutes an important energy sink.
 In fact, during the first million yr of cooling, the  energy lost  by neutrino
emission is of  about the same order of magnitude  as the gravothermal
energy  release, remaining larger  than the  star luminosity until the cooling time
reaches about $\log(t)\sim 7$.
As  the white  dwarf cools,  the temperature  of the
degenerate  core   decreases,  thus  neutrino  emission   ceases  and,
consequently,  the neutrino  luminosity abruptly  drops.
 It is during these stages that element diffusion strongly modifies
the internal chemical profiles. The  resulting chemical stratification will  be discussed
below.    
At $\log(t)\sim  8.3$ crystallization sets in at  the center of
the  white dwarf.  This  results in  the release  of  latent heat  and
gravitational  energy due  to  oxygen-neon  phase separation.   Note
that,  as   a  consequence   of  this   energy  release,   during  the
crystallization  phase  the  surface  luminosity is  larger  than  the
gravothermal  luminosity.   This phase  lasts  for  $2.5 \times  10^9$
years. Finally, at  $\log(t)  \sim 9$,  the  temperature of  the
crystallized core drops below the Debye temperature, and consequently,
the heat  capacity decreases.  Thus,  the white dwarf enters   the
so-called ``Debye cooling phase'', characterized by a rapid cooling.

\begin{figure}
\centering
\includegraphics[clip,width=\columnwidth]{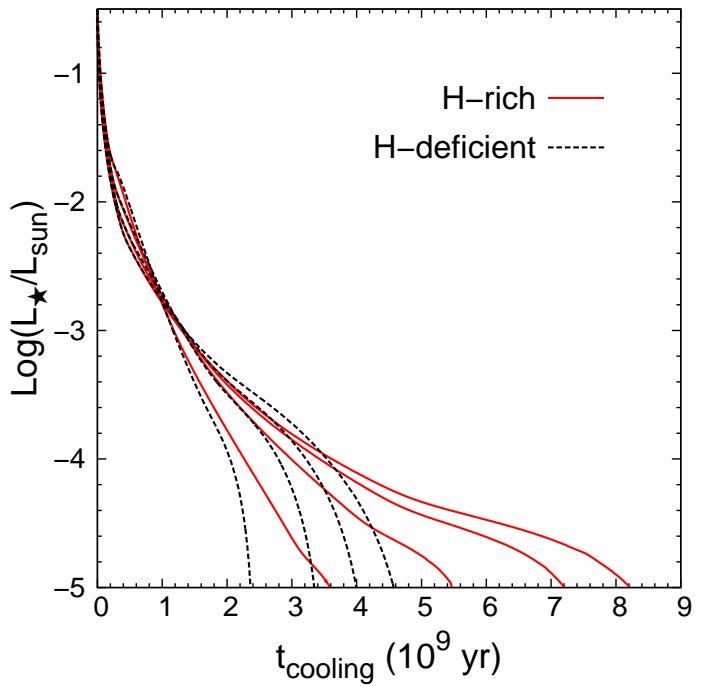}
\caption{Solid (dashed) lines display the cooling times for our hydrogen-rich (deficient) white dwarf sequences.
At low luminosities and from left to right,
stellar masses of both set of sequences are  $1.29 M_\sun, 1.23 M_\sun, 1.16 M_\sun$ and $1.10  M_\sun$.}
\label{Fig:5}
\end{figure}

The cooling times for all of our white dwarf sequences are displayed
in  Fig.~\ref{Fig:5}.    These cooling times are
also   listed  in   Table \ref{tabla1}  at   some  selected   stellar
luminosities.   { Our hydrogen-deficient} sequences have been 
calculated by considering recent advancement in the treatment of
energy transfer in dense helium atmospheres, \cite[see][for details]
{2017ApJ...839...11C, 2018MNRAS.473..457R}. As shown in \cite{2017ApJ...839...11C}, 
detailed non-gray model atmospheres
are needed to derive realistic cooling ages of cool,  helium-rich white dwarfs. 
 At   intermediate    luminosities,
hydrogen-deficient  white dwarfs  evolve  slightly  slower than  their
hydrogen-rich  counterparts. This result is in line with previous studies
of hydrogen-deficient white dwarfs \citep{2017ApJ...839...11C} and the
reason for this is that convective coupling  (and the
associated release  of internal energy) occurs  at higher luminosities
in hydrogen-deficient white dwarfs, with the consequent lengthening of
cooling times at those luminosities. By contrast, at
low-luminosities, hydrogen-deficient  white dwarfs evolve  markedly faster than
hydrogen-rich  white dwarfs.   This is due to the fact that, at  those stages,  the
thermal  energy  content of  the  hydrogen-deficient  white dwarfs  is
smaller, and   more importantly, because in  these white dwarfs, the  outer layers
are  more transparent  to radiation. Note in this sense that the
 $1.10 \, M_\sun$  hydrogen-rich sequence needs 8.2 Gyr to reach the lowest 
 luminosities, while the hydrogen-deficient sequence of the same mass evolves 
 in only 4.6 Gyr to the same luminosities. Note also the different cooling behavior
 with the stellar mass, particularly the fast cooling of the $1.29 \, M_\sun$ 
 hydrogen-rich sequence, our most massive sequence, which  reaches $\log(L_\star/L_\sun)= -5$ in 
 only 3.6 Gyr, which is even shorter (2.4 Gyr) in the case of the
 hydrogen-deficient counterpart. These short cooling times that characterize 
 the most massive sequences reflect that, at such
 stages, matter in  most of the white dwarf star has entered the Debye regime, with the 
 consequent strong reduction in the specific heat of ions \citep[see][for  details]{2010A&ARv..18..471A}.

{ All our hydrogen-deficient white dwarf sequences experience
carbon enrichment} in the outer layers as a result of convective mixing.
The   outer  convective  zone grows  inwards   and 
 when the luminosity of the star has decreased to $\log(L_\star/L_\sun) \sim -2.5$,
 it  penetrates into deeper layers where heavy elements such as
carbon  and  oxygen  are abundant.   Consequently,  convective  mixing
dredges  up   these  heavy  elements,  and   the  surface chemical composition
changes. In particular, the  surface layers are predominantly enriched
in carbon. These results are in line with the predictions of 
\cite{2017ApJ...839...11C} for hydrogen-deficient white dwarfs of intermediate mass.

\begin{table*}
\centering
\caption{Cooling times of our hydrogen-rich (HR) and hydrogen-deficient (HD) white dwarf sequences at selected luminosities.}
\begin{tabular}{ccccccccc}
\hline
\hline
$\log(L_\star/L_\sun)$ & \multicolumn{8}{c}{$t$ (Gyr)}\\
\hline
 & $1.10$(HR) & $1.16$(HR) & $1.23$(HR) & $1.29$ (HR) & $1.10$ (HD) & $1.16$ (HD) & $1.23$ (HD)& $1.29$ (HD)\\
\cline {2-9}
 $-2.0$ & $0.274$ & $0.290$ & $0.356$ & $0.437$ & $0.266$ & $0.289$ & $0.361$ & $0.479$\\
 $-3.0$ & $1.318$ & $1.310$ & $1.320$ & $1.185$ & $1.367$ & $1.354$ & $1.325$ & $1.173$\\
 $-3.5$ & $2.236$ & $2.173$ & $2.043$ & $1.692$ & $2.457$ & $2.268$ & $2.010$ & $1.590$\\
 $-4.0$ & $3.625$ & $3.427$ & $2.999$ & $2.265$ & $3.547$ & $3.217$ & $2.793$ & $2.048$\\
 $-4.5$ & $6.203$ & $5.390$ & $4.132$ & $2.876$ & $4.209$ & $3.739$ & $3.171$ & $2.273$\\
 $-5.0$ & $8.225$ & $7.213$ & $5.467$ & $3.594$ & $4.580$ & $3.996$ & $3.346$ & $2.362$\\
\hline
\hline
\end{tabular}
\label{tabla1}
\end{table*}

\begin{table*}                                                                  
\centering                                                                     
\caption{Stellar mass (solar mass) and cooling ages (Gyr) as 
predicted by our sequences under the assumption that they harbour O-Ne cores for selected ultra-massive white dwarfs in the literature. The letter "V" (variable) 
indicates that the star is a ZZ Ceti star. The last column gives the references from which the $T_{\rm eff}$ and $\log g$ values have been extracted.}                                              
\begin{tabular}{lccrccc}                                                   
\hline                                                                     
\hline                                                                     
Star & Spectral Type& $\log(g)({\rm cgs})$& $T_{\rm eff} ({\rm K})$& $M_\star/ M_\sun$ & $t ({\rm Gyr})$ & {\rm Reference}\\  
\hline                                                                     
  SDSS J\,090549.46$+$134507.87 & DA &    8.875 &    6774   &    1.110  & 3.966 & \cite{2016MNRAS.455.3413K} \\
  SDSS J\,000901.20$+$202606.80 & DA &    8.857 &   11081   &    1.104  & 1.706 & "\\
  SDSS J\,002113.16$+$192433.62 & DA &    8.920 &   11555   &    1.134  & 1.655 & "\\
  SDSS J\,003608.73$+$180951.52 & DA &    9.250 &   10635   &    1.248  & 2.121 & "\\
  SDSS J\,005142.50$+$200208.66 & DA &    9.080 &   14593   &    1.197  & 1.244 & "\\
  SDSS J\,013853.19$+$283207.13 & DA &    9.402 &    9385   &    1.288  & 2.305 & "\\ 
  SDSS J\,015425.78$+$284947.71 & DA &    8.959 &   11768   &    1.153  & 1.652 & "\\
  SDSS J\,001459.15$+$253616.37 & DA &    8.812 &   10051   &    1.081  & 1.982 & "\\
  SDSS J\,004806.14$+$254703.56 & DA &    8.885 &    9388   &    1.116  & 2.322 & "\\
  SDSS J\,005122.96$+$241801.15 & DA &    9.170 &   10976   &    1.226  & 2.069 & "\\
  SDSS J\,224517.61$+$255043.70 & DA &    8.990 &   11570   &    1.165  & 1.734 & "\\
  SDSS J\,222720.65$+$240601.31 & DA &    8.947 &    9921   &    1.146  & 2.190 & "\\
  SDSS J\,232257.27$+$252807.42 & DA &    8.882 &    6190   &    1.113  & 4.581 & "\\
  SDSS J\,164642.67$+$483207.96 & DA &    8.999 &   15324   &    1.169  & 1.042 & "\\
  SDSS J\,110054.91$+$230604.01 & DA &    9.470 &   11694   &    1.307  & 1.828 & "\\
  SDSS J\,111544.64$+$294249.50 & DA &    9.136 &    8837   &    1.214  & 2.770 & "\\
  SDSS J\,102720.47$+$285746.16 & DA &    9.053 &    8874   &    1.186  & 2.713 & "\\
  SDSS J\,100944.29$+$302102.03 & DA &    9.161 &    6639   &    1.222  & 3.893 & "\\
  SDSS J\,130846.79$+$424119.60 & DA &    8.970 &    7237   &    1.156  & 3.668 & "\\
  SDSS J\,101907.08$+$484805.90 & DA &    9.231 &   12582   &    1.243  & 1.691 & "\\
  SDSS J\,122943.28$+$493451.45 & DA &    9.240 &   16889   &    1.246  & 1.083 & "\\
  SDSS J\,110510.71$+$474804.08 & DA &    9.089 &    9538   &    1.198  & 2.460 & "\\
  SDSS J\,150417.23$+$553900.45 & DO &    9.267 &    6360   &    1.244  & 2.929 & "\\
  SDSS J\,145009.87$+$510705.21 & DA &    9.180 &   11845   &    1.229  & 1.849 & "\\
  SDSS J\,132208.52$+$551939.16 & DAH&    9.098 &   17136   &    1.204  & 0.939 & "\\
  SDSS J\,004825.11$+$350527.94 & DA &    8.887 &    7516   &    1.116  & 3.367 & "\\
  SDSS J\,013550.03$-$042354.59 & DA &    9.150 &   12651   &    1.220  & 1.659 & "\\
  SDSS J\,102553.68$+$622929.41 & DAH&    9.356 &    9380   &    1.276  & 2.359 & "\\
  SDSS J\,104827.74$+$563952.68 & DA &    8.829 &    9680   &    1.090  & 2.134 & "\\
  SDSS J\,112322.47$+$602940.06 & DA &    8.845 &   13611   &    1.099  & 1.121 & "\\
  SDSS J\,110036.93$+$665949.42 & DA &    9.383 &   22251   &    1.286  & 0.760 & "\\
  SDSS J\,004920.03$-$080141.71 & DA &    9.403 &   11648   &    1.289  & 1.849 & "\\
  SDSS J\,013514.18$+$200121.97 & DA &    9.370 &   17134   &    1.281  & 1.130 & "\\
  SDSS J\,093710.25$+$511935.12 & DA &    8.969 &    7030   &    1.155  & 3.827 & "\\
  SDSS J\,234929.60$+$185119.52 & DA &    8.935 &    6966   &    1.139  & 3.848 & "\\   
  SDSS J\,232512.08$+$154751.27 & DA &    9.063 &   10083   &    1.190  & 2.234 & "\\  
  SDSS J\,234044.83$+$091625.96 & DA &    9.234 &    6166   &    1.242  & 3.957 & "\\  
  SDSS J\,003652.69$+$291229.48 & DA &    9.070 &   10284   &    1.192  & 2.182 & "\\
  SDSS J\,000011.57$-$085008.4  & DQ &    9.230 &   10112   &    1.236  & 2.299 & \cite{2013ApJS..204....5K}\\    
  SDSS J\,000052.44$-$002610.5  & DQ &    9.320 &   10088   &    1.257  & 2.192 & "\\   
  GD50 (WD 0346$-$011)          & DA &    9.200 &   42700   &    1.241  & 0.064 & \cite{2011ApJ...743..138G}\\
 GD518 (WD J165915.11+661033.3) {\rm(V)}   &DA& 9.080&  12030  & 1.196  & 1.719 & " \\  
  SDSS J\,072724.66$+$403622.0  & DA &    9.010 &   12350   &    1.172  & 1.573 & \cite{2017MNRAS.468..239C}\\     
  SDSS J\,084021.23$+$522217.4 {\rm (V)} & DA &  8.930 & 12160  &1.139  & 1.523 & "\\  
  SDSS J\,165538.93$+$253346.0  & DA &    9.200 &   11060   &    1.234  & 2.035 & "\\ 
  SDSS J\,005047.61$-$002517.1  & DA &    8.980 &   11490   &    1.162  & 1.744 & \cite{2004ApJ...607..982M} \\ 
 BPM 37093 (LTT 4816)  {\rm(V)}  & DA &    8.843 &   11370  &     1.097  & 1.608 & \cite{2016IAUFM..29B.493N}\\    
\hline                                                                     
\hline                                                                     
\end{tabular}                                                                
\label{tabla2}                                                                 
\end{table*}

\begin{table}
\centering
\caption{Percentages of crystallized mass of our hydrogen-rich sequences and effective temperature at which they occur.}
\begin{tabular}{ccccc}
\hline
\hline
Crystallized mass & \multicolumn{4}{c}{$\log(T_{\rm eff})$ (K)}\\
\hline
 & $1.10 M_\sun$ & $1.16  M_\sun$ & $1.23 M_\sun$ & $1.29 M_\sun$\\
\cline {2-5}
$0 \%$ & $4.31$ & $4.38$ & $4.46$ & $4.58$\\
$20 \%$ & $4.26$ & $4.32$ & $4.41$ & $4.54$\\
$40 \%$ & $4.22$ & $4.29$ & $4.38$ & $4.51$\\
$60 \%$ & $4.17$ & $4.23$ & $4.34$ & $4.46$\\
$80 \%$ & $4.09$ & $4.16$ & $4.26$ & $4.39$\\
$90 \%$ & $4.03$ & $4.10$ & $4.20$ & $4.33$\\
$95 \%$ & $3.91$ & $3.95$ & $4.10$ & $4.25$\\
$99 \%$ & $3.77$ & $3.83$ & $4.02$ & $4.10$\\
\hline
\hline
\end{tabular}
\label{tabla3}
\end{table}

\begin{figure*}
\centering
\includegraphics[clip,width=2\columnwidth]{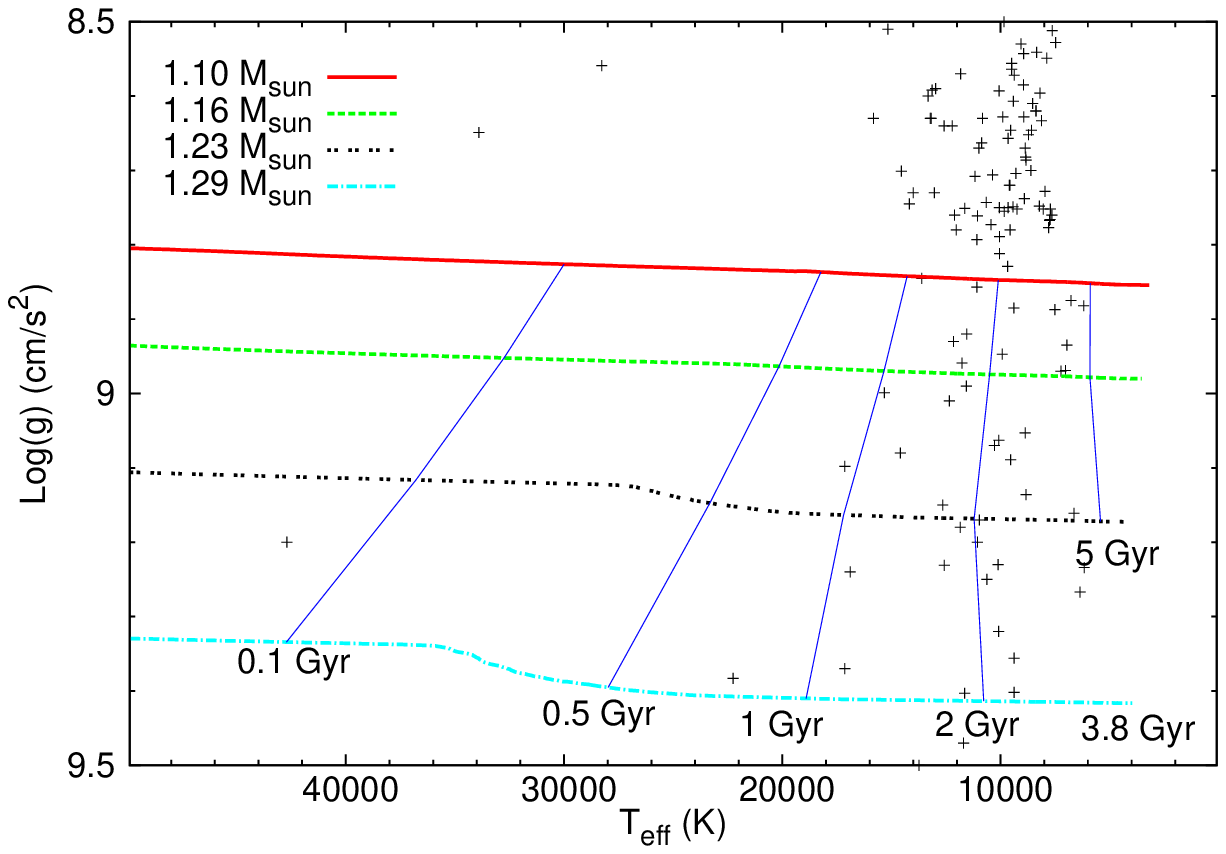}
\caption{The hydrogen-rich sequences in the plane $\log(g)-T_{\rm eff}$. Blue solid lines  
display isochrones of 0.1, 0.5, 1, 2, and 5 Gyr. Crosses indicate the location of 
ultra-massive white dwarfs from \cite{2004ApJ...607..982M,2016IAUFM..29B.493N,2011ApJ...743..138G,2013ApJS..204....5K,2015MNRAS.450.3966B,2016MNRAS.455.3413K,2017MNRAS.468..239C}.}
\label{Fig:6}
\end{figure*}

The evolution of our ultra-massive white dwarf sequences in   the plane $\log(g)-T_{\rm eff}$ is depicted in  Fig. \ref{Fig:6} 
together with observational expectations taken from \cite{2004ApJ...607..982M,2016IAUFM..29B.493N,2011ApJ...743..138G,2013ApJS..204....5K,2015MNRAS.450.3966B,2016MNRAS.455.3413K,2017MNRAS.468..239C}. 
In addition, isochrones of 0.1, 0.5,  1, 2, and 5  Gyr connecting the curves are shown. 
For these white dwarfs, we estimate from our sequences the stellar mass and cooling age (we elect those for which their surface gravities are larger than 8.8).
Results are shown in Table \ref{tabla2}. Note that for most of the observed white dwarfs, the resulting cooling age is in the range $1-4$ Gyr, and many of them have 
stellar masses  above $1.25 M_\sun$. Note also from Fig. \ref{Fig:6} the change of slope of the isochrones, reflecting the well known dependence of cooling times 
on the mass of the white dwarf, i.e, at early stages, evolution proceeds slower in more massive white dwarfs, while the opposite trend is found at advanced stages.

\begin{figure}
\centering
\includegraphics[clip,width=\columnwidth]{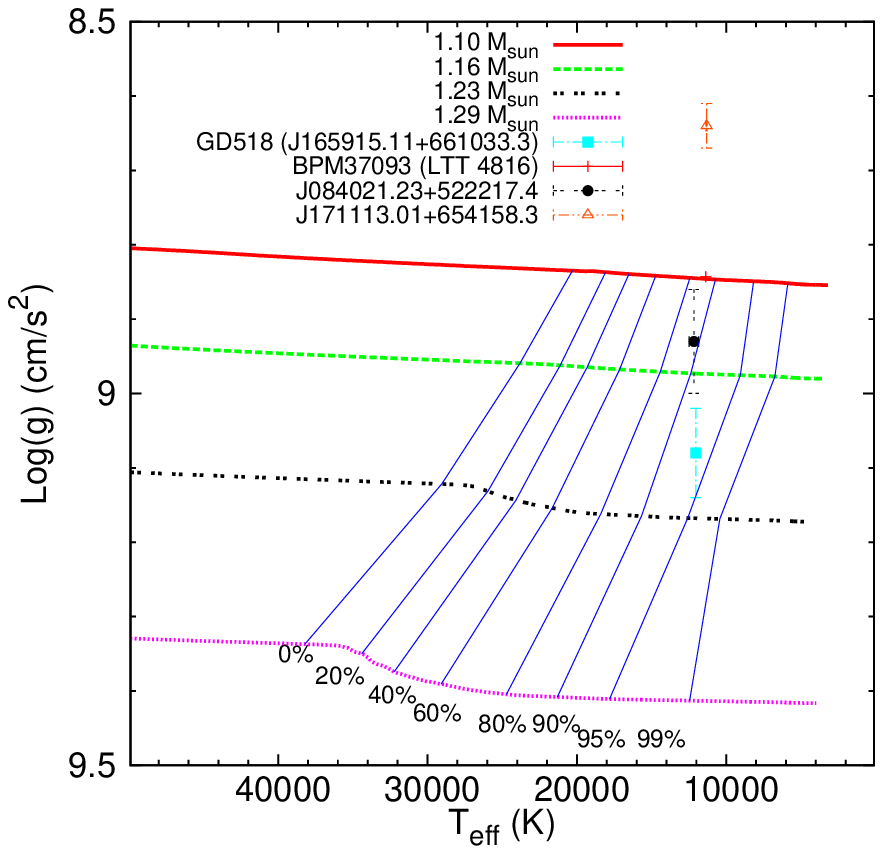}
\caption{The hydrogen-rich sequences in the plane $\log(g)-T_{\rm eff}$. Blue solid lines  
indicate 0, 20, 40, 60, 80, 90, 95 and 99\% of crystallized mass. 
The symbols with error bars indicate the location of the known pulsating ultra-massive DA white
dwarfs \citep{2004ApJ...607..982M,2013ApJ...771L...2H, 2017MNRAS.468..239C,2016IAUFM..29B.493N}.}
\label{Fig:7}
\end{figure}

In Fig. \ref{Fig:7} we display our hydrogen-rich sequences in the plane
$\log(g)-T_{\rm eff}$ together with observational expectations for pulsating
massive white dwarfs taken from \cite{2004ApJ...607..982M,2013ApJ...771L...2H, 2017MNRAS.468..239C,2016IAUFM..29B.493N}. 
Also with  solid lines we show the 
0, 20, 40 ,60, 80, 90, 95 and 99 \% of the crystallized mass of the star. 
Note that all of the observed pulsating white dwarfs with masses larger than $1.1 M_\sun$
fall in the region where more than 80\% of their mass is expected to be crystallized. 
It is expected, as we will discuss in a forthcoming paper, that crystallization process 
affects the pulsation properties of massive ZZ Ceti stars, { as it has also been shown 
by \cite{1999ApJ...526..976M,2004A&A...427..923C,2005A&A...429..277C,2005ApJ...622..572B}. }

The effective temperature at various percentage of crystallized mass is also listed in Table \ref{tabla3}.
{ Note that, at the onset of crystallization, the highest mass} 
sequences exhibit a marked increase in their
surface gravities. This behavior is a consequence of the change in the chemical abundances 
of $^{16}$O and $^{20}$Ne
during the crystallization. As the abundance of $^{20}$Ne grows in the
inner regions of the white dwarf, its radius 
decreases, and consequently its surface gravity increases. 
{Crystallization sets in at similar luminosities and
effective temperatures in a hydrogen-deficient as in a hydrogen-rich white dwarf with the same mass.}
Hydrogen-deficient cooling sequences
are not shown in this Figure since they 
exhibit a similar behavior but their surface gravities are slightly larger, since their radius are relatively smaller. 

\begin{figure}
\centering
\includegraphics[clip,width=\columnwidth]{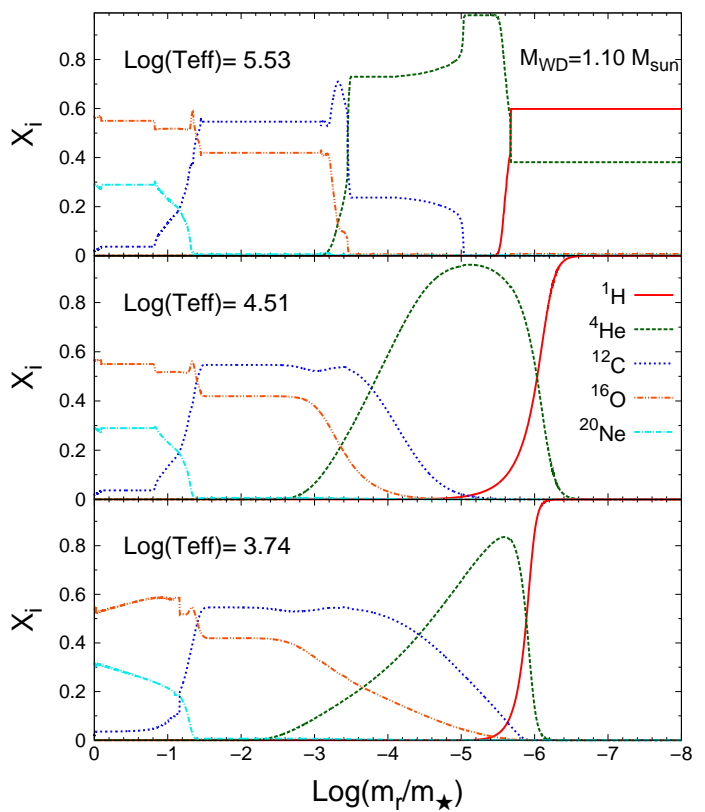}
\caption{Inner abundance distribution for $1.10 M_\sun$ hydrogen-rich models at three selected effective temperatures, as indicated.}
\label{Fig:8}
\end{figure}

\begin{figure}
\centering
\includegraphics[clip,width=\columnwidth]{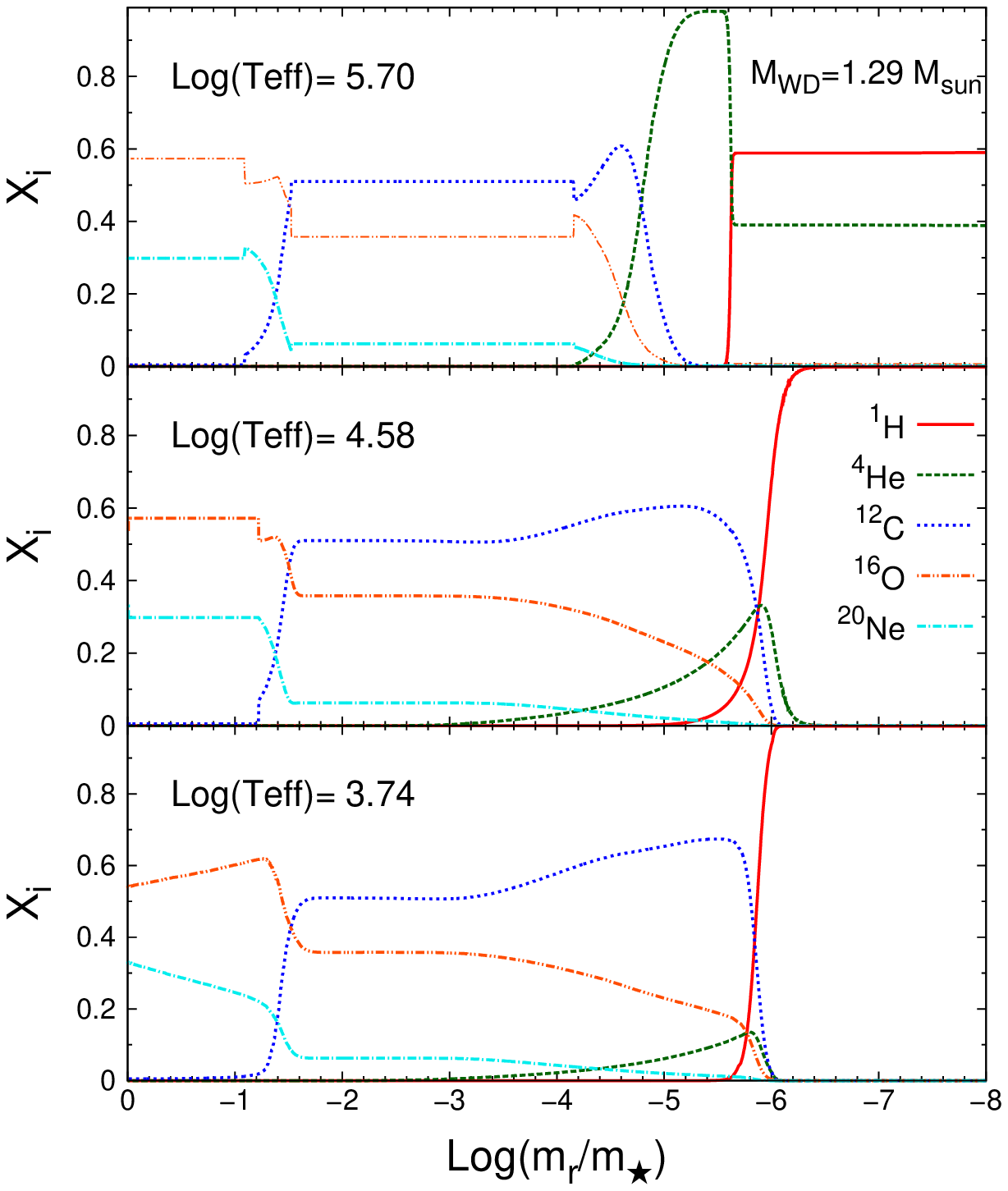}
\caption{Same as Fig. \ref{Fig:8} but for $1.29 M_\sun$ models.}
\label{Fig:9}
\end{figure}

{ Element diffusion profoundly alters the inner abundance distribution from the early cooling stages of our massive white dwarf models.} This is borne out by Figs. \ref{Fig:8} and \ref{Fig:9}, which display the abundance distribution 
in the whole star at three selected effective temperatures for the 1.10 and $1.29 M_\sun$ hydrogen-rich white dwarf models, respectively. 
Note that, as a result
of gravitational settling, all heavy elements are depleted from the outer layers.
{
Note also that that initial chemical discontinuities
 are
strongly smoothed out.} But more importantly, the initial helium and carbon distribution
in the deep envelope result markedly changed, 
particularly in the most massive models, where the initial pure helium buffer has almost
vanished when evolution has reached low effective temperatures. This is quite different from the
situation encountered in white dwarfs of intermediate mass. These changes in the 
helium and carbon profiles 
affects the radiative opacity in the envelope and thus the cooling times at late stages.

\begin{figure}
\centering
\includegraphics[clip,width=\columnwidth]{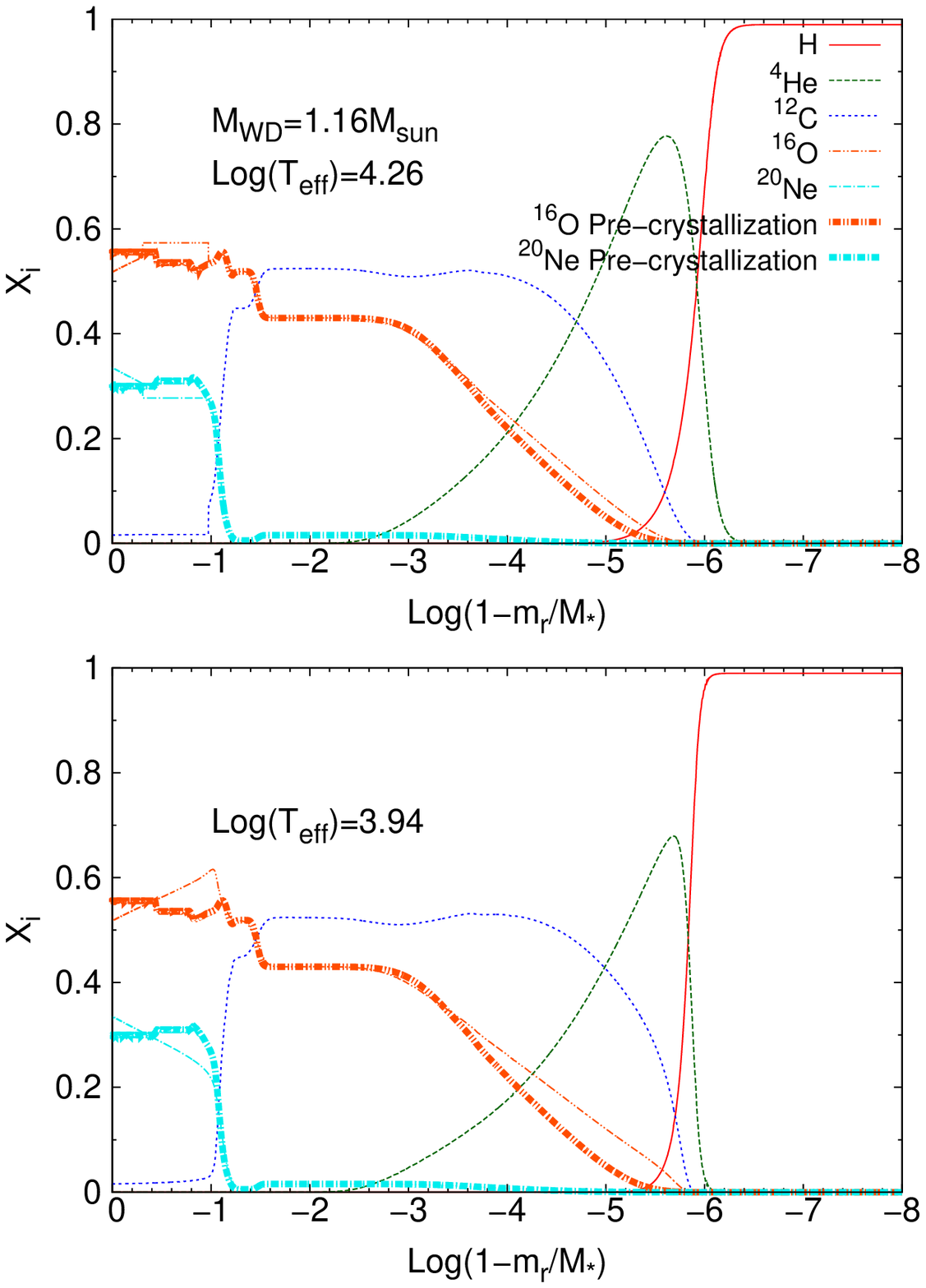}
\caption{ Change in the chemical profiles of our $1.16M_\sun$ hydrogen-rich white 
dwarf model induced by
the phase separation process during crystallization. 
Top (bottom) panel depicts the chemical profile at  $\log(T_{\rm eff})= 4.26 (3.94)$.
For comparison, the abundances of $^{16}$O and $^{20}$Ne right before phase separation are also plotted with thick lines
in both panels.}
\label{Fig:10}
\end{figure}

The other physical process that changes the core chemical distribution 
during white dwarf evolution is, as we mentioned, phase separation during  crystallization. The imprints of phase
separation on the core chemical composition can be appreciated in the bottom panels of Fig.  \ref{Fig:8} and \ref{Fig:9},
and more clearly in Fig. \ref{Fig:10}, which illustrates the change in the abundances of $^{20}$Ne and
 $^{16}$O in a  $1.16  M_\sun$ model shortly after the occurrence of crystallization (top panel)
 and  by the time a large
 portion of the star has crystallized (bottom panel). The chemical abundances of $^{20}$Ne and 
 $^{16}$O right before the crystallization
sets in, are plotted with thick dashed lines.
For this stellar mass, crystallization starts at the center of the
{ star at $\log(L_\star/L_\sun)
\sim -1.8$. 
Note that, in the top panel, (the crystallization front is at $\log(M_r/M_\star) \sim -0.4$) the initial $^{20}$Ne and  $^{16}$O abundances have
strongly been changed by the process of phase separation and the induced
mixing in the fluid layers above the core, which extends upwards to
$\log(M_r/M_\star) \sim -1$. Other elements apart from  $^{16}$O and  $^{22}$Ne
are not taken into account in the phase separation process, and the slight change
shown in their abundances is due only to element diffusion. 

To properly 
assess the phase separation process during crystallization, it should be necessary to
consider a 5-component crystallizing plasma composed in our case by  $^{12}$C ,  $^{16}$O,  $^{20}$Ne,$^{23}$Na and
 $^{24}$Mg, which are the most abundant elements in
 the white dwarf core (see Figure \ref{Fig:profiles2}). 
Such 5-component phase diagram is not available
in the literature. However, Prof. A. Cumming has provided us with  the final abundances in the solid phase in the center of the 
$1.10\, \rm M_\odot$ white dwarf model, considering a given 5-component composition \footnote{A. Cumming, personal communication.}. The abundances of $^{12}$C ,  $^{16}$O, 
$^{20}$Ne,$^{23}$Na and
 $^{24}$Mg at the center of this model
 right before crystallization occurs
 are listed in Table \ref{tablacumming}, together
with the final abundances in the solid phase predicted by the 5-component calculations,
and those predicted by the phase diagram for a $^{16}$O-$^{20}$Ne mixture shown
in Figure \ref{Fig:PD}. The abundances of $^{16}$O and $^{20}$Ne are noticeably altered
by crystallization
regardless of the treatment considered. However, considering a 2-component phase diagram results
in a stronger phase separation of $^{16}$O and $^{20}$Ne.
Nevertheless, in this treatment the abundances of trace elements
 $^{12}$C, $^{23}$Na and
 $^{24}$Mg are not altered by the crystallization process. 
 The sum of the abundances of these trace elements is lower than 
 15\% in the core of all our ultra-massive white dwarf models and
 we do not expect this to alter substantially the evolutionary timescales. 
 To properly assess the effects of considering a 5-component phase diagram
 on the cooling times of white dwarfs 
 it should be necessary calculate the 
 evolution of the white dwarf model through the entire crystallization process, for which we would
 require the full phase diagram, not available at the moment
 of this study.}

\begin{table}
\centering
\caption{ Abundances at the center of the $1.10\, \rm M_\odot$ white dwarf
model before crystallization, and the final abundances in the solid phase
resulting of considering 
a 5-component mixture of $^{12}$C ,  $^{16}$O, 
$^{20}$Ne,$^{23}$Na and
 $^{24}$Mg, and the 2-component $^{16}$O-$^{20}$Ne phase diagram shown in Figure \ref{Fig:PD}.}
\begin{tabular}{cccc}
\hline
\hline
& Initial & Solid 5-component & Solid 2-component\\
\hline
 $^{12}$C  & $0.0167$  & $0.0082$ & $0.0167$\\
 $^{16}$O & $0.5624$  & $0.5561$ & $0.5450$\\
 $^{20}$Ne & $0.2921$  & $0.3289$ & $0.3311$\\
 $^{23}$Na & $0.0538$ & $0.0579$  & $0.0538$\\
 $^{24}$Mg & $0.0513$  & $0.0489$ & $0.0513$\\
\hline
\hline
\end{tabular}
\label{tablacumming}
\end{table}


\begin{figure}
\centering
\includegraphics[clip,width=\columnwidth]{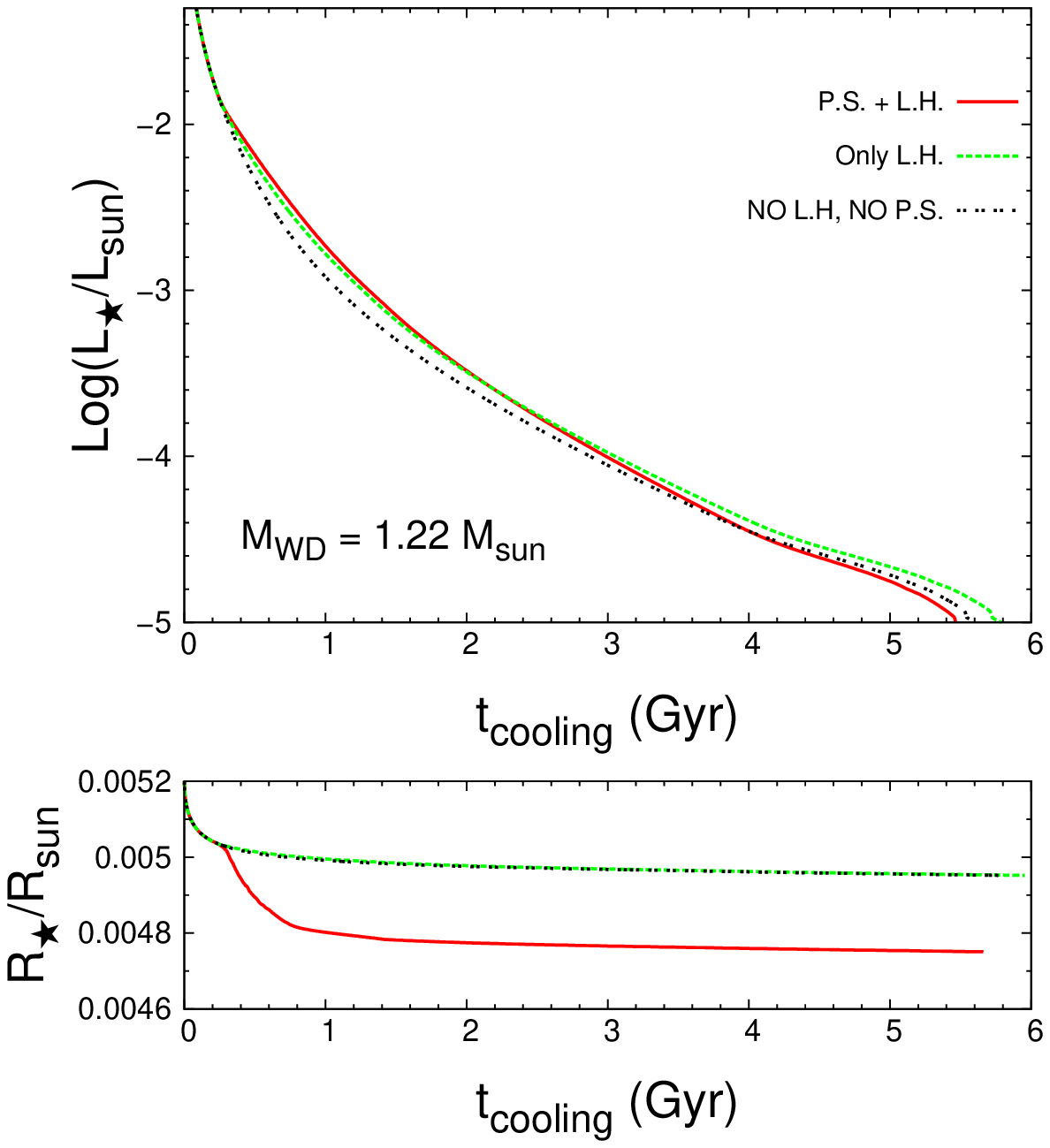}
\caption{Top panel: Cooling times of our $1.22 M_\sun$ hydrogen-rich sequence 
when crystallization is neglected (double-dotted line), when only latent heat 
is considered during crystallization (dotted line), and when both latent heat and energy from phase separation are
considered during crystallization (solid line).
Bottom panel: White dwarf radius in terms of the cooling time for these evolutionary tracks.}
\label{Fig:11}
\end{figure}

The phase separation process of $^{20}$Ne and
$^{16}$O releases appreciable energy, see Fig. \ref{Fig:4},
so as to impact the white dwarf cooling times. This can be seen 
in Fig. \ref{Fig:11}, which shows the cooling times for 
our  $1.22 M_\sun$ hydrogen-rich sequence (upper
panel)  when crystallization is neglected (double-dotted line), w
hen only latent heat is considered during
crystallization (dotted line), and when both latent heat and energy from phase separation are
considered during crystallization (solid line). Clearly, the energy
resulting from crystallization, in particular the release of latent heat, 
increases substantially the cooling times of the ultra-massive white dwarfs. 
The inclusion of energy from  phase separation leads to an additional 
delay on the cooling times (admittedly less than the delay caused by latent heat)
at intermediate luminosities. But below  $\log(L_\star/L_\sun)\sim -3.6$, when most of the star has
crystallized, 
phase separation accelerates the cooling times. At these stages, no more energy
is delivered by phase separation, but the changes in the chemical
profile induced by phase separation have strongly altered both the structure and thermal
properties of the cool white dwarfs, 
impacting their rate of cooling. Note in this sense, the change in the radius of the white dwarf 
that results from the inclusion of
phase separation (bottom panel of
Fig. \ref{Fig:11}). In fact, the star radius becomes smaller due 
to the increase of neon in the core during crystallization. 
As we mentioned, this explains the increase of the surface gravity of 
our sequences in the case of
phase separation is considered, see Figs. \ref{Fig:6} and \ref{Fig:7}

\begin{figure}
\centering
\includegraphics[clip,width=\columnwidth]{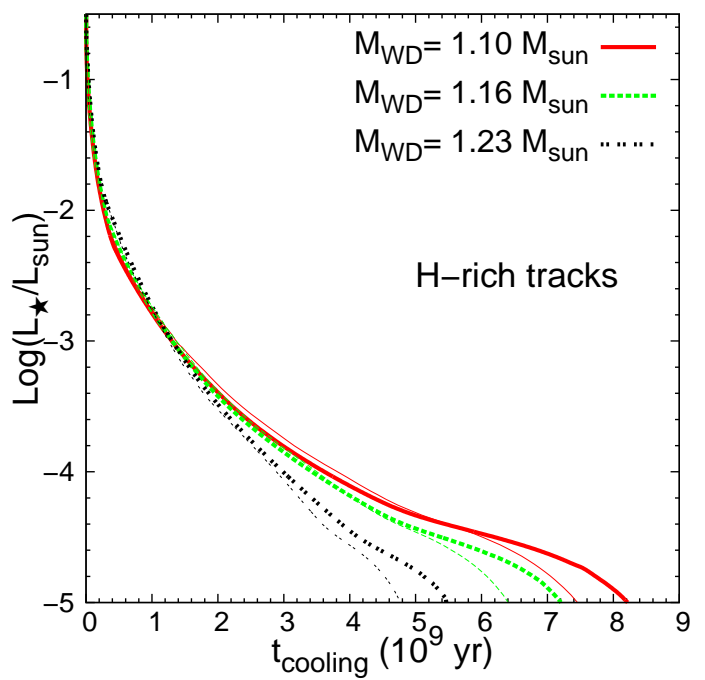}
\caption{Cooling times of our hydrogen-rich white dwarf sequences with 
$1.10 M_\sun$, $1.16 M_\sun$ and $1.23 M_\sun$ 
 (thick lines), as compared with the cooling sequences of A07 of similar masses (thin lines).}
\label{Fig:12}
\end{figure}

\begin{figure}
\centering
\includegraphics[clip,width=\columnwidth]{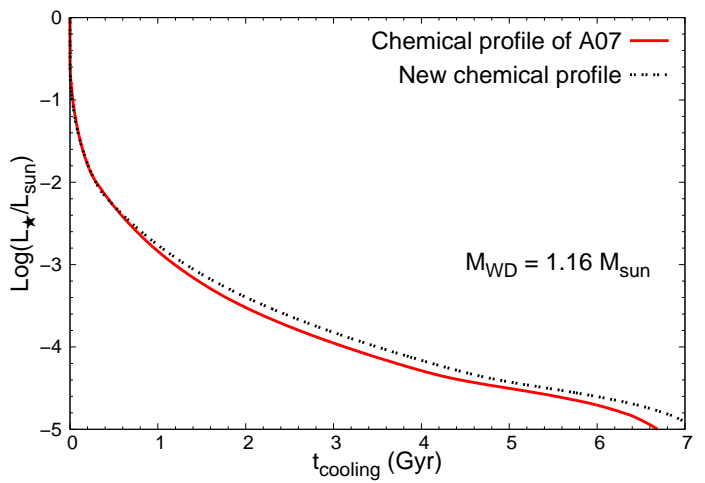}
\caption{Cooling times of  $1.16  M_\sun$ hydrogen-rich white dwarf models
without phase separation resulting from the use of different chemical profiles. 
Solid red line corresponds to the cooling sequence using
our current stellar evolutionary code but implanting  the chemical profile considered
in A07. Black dotted line corresponds to the cooling sequence calculated using
our new chemical profile (plotted in the top right panel of Fig. \ref{Fig:profiles2}).}
\label{Fig:13}
\end{figure}

The present evolutionary sequences of ultra-massive white dwarfs constitute an improvement 
over those presented in A07. The comparison between the evolutionary 
sequences of both studies is presented in Fig. \ref{Fig:12} for the $1.10 M_\sun$, $1.16 M_\sun$
and $1.23 M_\sun$ hydrogen-rich sequences. Note that
appreciable differences in the cooling times exist between both set of sequences. In particular, 
the present calculations predict shorter ages at 
intermediate luminosities, but this trend is reversed
at very low surface luminosities, where our new sequences evolve markedly slower than in A07.

To close the paper, we attempt to trace back the origin of such differences. We begin by examining the impact of the new chemical profiles,
as compared with that used in A07 (as illustrated in Figure 4 of  \cite{2004A&A...427..923C}) which is the same used for all white dwarf sequences in A07.
To this end,  we have computed two artificial 
white dwarf sequences by neglecting
phase separation during crystallization.
Comparison is made in Fig. \ref{Fig:13}, which shows the cooling times of a $1.16  M_\sun$ 
hydrogen-rich white dwarf model resulting from the use of
the chemical profile considered in  A07 (solid line) and the chemical profile
employed in the current study (dotted line).  Note that 
the use of new chemical profiles employed in the present study \citep{2010A&A...512A..10S}
predicts larger cooling times than the use of the chemical profiles of   \cite{1997ApJ...485..765G} 
considered in A07. This is due to not only to the 
different core chemical stratification in both cases but also to the different predictions for the
helium buffer mass expected in the white dwarf envelopes,  which affects the cooling rate of cool
white dwarfs. In this sense, 
the full computation of evolution of progenitor stars along the thermally pulsing SAGB constitutes
an essential aspect that cannot 
be overlooked in any study of the cooling of massive white dwarfs.

\begin{figure*}
\centering
\includegraphics[clip,width=2\columnwidth]{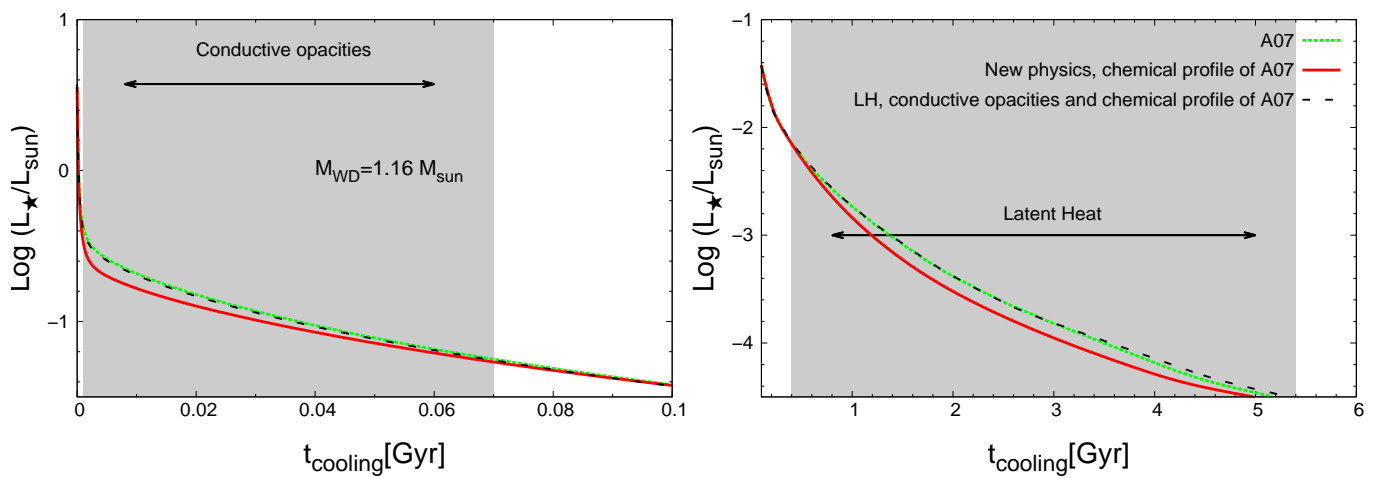}
\caption{Cooling times of  $1.16  M_\sun$ hydrogen-rich white dwarf models without phase separation and considering the same chemical profiles of A07, but adopting different microphysics. The cooling sequence
of A07 is displayed using a green dotted line, which is based on
an old treatment of conductive opacity and latent heat. The red solid line shows  the cooling times calculated using our current numerical code. The dashed  line shows prediction given by
our current numerical code but  with old microphysics, this is, the same conductive opacities and treatment of latent heat as 
considered in A07. 
The left panel amplifies the early stages of the white dwarf stage. The right panel shows the rest of the
white dwarf cooling track.}
\label{Fig:14}
\end{figure*}

Improvements in the microphysics considered in the computation of our new sequences also 
impact markedly the cooling times;
this is particularly true regarding the treatment of conductive opacities and the release 
of latent heat during crystallization. 
Specifically, in the present sequences we make use of the conductive opacity as given
in \cite{2007ApJ...661.1094C}, in contrast to
A07 where the older conductive opacities of \cite{1994ApJ...436..418I} were employed. 
The resulting impact on the cooling time becomes 
apparent from Fig. \ref{Fig:14}. Here we compare the cooling times for
$1.16 M_\sun$ white dwarf white dwarf models having the same chemical composition as 
in A07 but adopting different microphysics. 
A close inspection of this figure reveals that the improvement in the microphysics 
considered in our current version of {\tt LPCODE} 
as compared with that used in A07, particularly the conductive opacity at intermediate 
luminosity and the treatment of latent heat  
during the crystallization phase at lower luminosities, lead to shorter cooling times. 
Note that when we use the old microphysics 
(and the same chemical profile) we recover the results of A07.

We conclude from Figs. \ref{Fig:13} and
\ref{Fig:14} that the  inclusion of detailed chemical profiles appropriate 
for massive white dwarfs resulting from SAGB progenitors and improvements
in the microphysics results in evolutionary sequences for these white dwarfs 
much more realistic than those presented in A07. 
These improvements together with the consideration of the effects of 
phase separation of $^{20}$Ne and $^{16}$O during crystallization 
yield accurate cooling times  for ultra-massive white dwarfs.

\begin{figure}
\centering
\includegraphics[clip,width=\columnwidth]{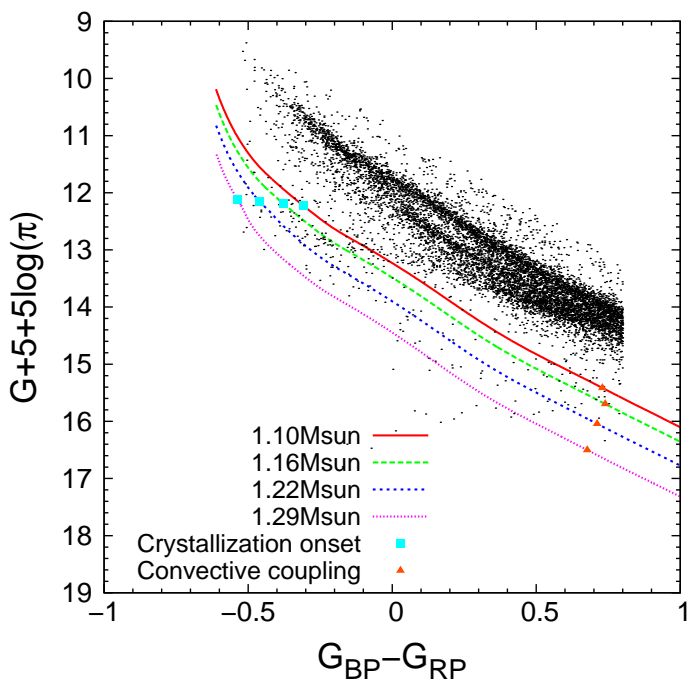}
\caption{ H-rich white dwarf cooling sequences in the color-magnitude diagram in GAIA bands,
together with the sample of white dwarfs within 100 pc, obtained by \cite{2018MNRAS.480.4505J}.
The filled squares indicate the moment when crystallization begins in each white dwarf cooling
sequence and the filled triangles indicate the moment
when convective coupling occurs.}
\label{Fig:15}
\end{figure}

{ Finally, we present our ultra-massive white dwarf cooling tracks
in GAIA photometry bands:  G, $\rm G_{BP}$ and $\rm G_{RP}$. These magnitudes 
have been obtained using detailed model atmospheres for H-composition described
in \cite{2012A&A...546A.119R}. The cooling tracks are plotted in the color-magnitude
diagram in Figure \ref{Fig:15}, together with the local sample of white
dwarfs within 100 pc from our sun of \cite{2018MNRAS.480.4505J}, 
in the color range:
$-0.52<(\rm G_{BP}- G_{RP})<0.80$.
The onset of
crystallization in our cooling sequences is indicated with filled squares. Note that crystallization occurs at approximately the same magnitude, $G+5+5log(\pi)\sim 12$. The moment when convective coupling is occurring 
in each white dwarf sequence is also indicated using filled triangles. Clearly, our ultra-massive white dwarf cooling
tracks fall below the vast majority of the white dwarf sample. The reason for this
relies on the mass distribution of the white dwarf sample, that exhibits a sharp peak
around $0.6\,\rm M_\odot$ \citep{2019MNRAS.482.5222T}. Thus, the vast majority of white
dwarfs will be characterized by larger luminosities than the ones present in
our ultra-massive white dwarfs. However,
a detailed analysis of this color-magnitude diagram
is beyond the scope of the present paper and we simply present white dwarf colors
for our ultra-massive white dwarfs, which are available for downloading.}

\section{Summary and conclusions}
\label{conclusions}
 
In this paper we have studied the evolutionary properties of ultra-massive white dwarfs
with $^{16}$O and $^{20}$Ne cores. 
For this purpose, we have calculated hydrogen-rich and hydrogen-deficient
white dwarf cooling sequences of $1.10, 1.16, 1.23$ and $1.29 M_\sun$,
resulting from solar metallicity progenitors with the help of
{\tt LPCODE} evolutionary code. 
These cooling sequences are appropriate for
the study of the massive white dwarf population in the solar neighborhood 
resulting from single evolution of progenitor stars. 
In our study we have considered 
initial chemical profiles for each white dwarf model  consistent 
with predictions of the progenitor evolution with
stellar masses in the range $9.0\leq M_{\rm ZAMS}/ M_\sun \leq 10.5$, as calculated in \cite{2010A&A...512A..10S}.
These chemical profiles are the result of the computation of full evolutionary sequences from the ZAMS, 
through the core hydrogen burning, 
core helium  burning, and the semidegenerate carbon burning during the  thermally-pulsing SAGB phase. Hence, not only a
realistic O-Ne inner profile is considered for each
white dwarf mass, but also realistic chemical profiles and intershell masses built up
during the SAGB are taken into account. In particular,  
the evolution through the entire SAGB phase provides us 
with  realistic values of the 
total helium content necessary to compute realistic cooling times
at low luminosities. 
We have calculated both hydrogen-rich and hydrogen-deficient white dwarf evolutionary sequences.  In particular our hydrogen-deficient sequences have been 
calculated by considering recent advancement in the treatment of
energy transfer in dense helium atmospheres.  Each evolutionary sequence was computed from the beginning of the cooling track at high luminosities down to the development of
the full Debye cooling at very low surface
luminosities, $\log(L_\star/L_\sun)= -5.5$. { We also provide
colors in the GAIA photometric bands for these white dwarf evolutionary sequences on the basis of models atmospheres of
\cite{2012A&A...546A.119R}.}

A relevant aspect of our sequences is that we have included
the release of energy and the ensuing
core chemical redistribution resulting from the phase 
separation of $^{16}$O and $^{20}$Ne induced by the crystallization. 
This constitutes a major improvement as compared with previous 
studies on the subject, like those of A07 and  \cite{2004A&A...427..923C}.
To this end, we incorporate the phase diagram of \cite{2010PhRvE..81c6107M}
suitable for $^{16}$O and $^{20}$Ne plasma, 
which provides us also with the correct temperature of crystallization. In addition, 
our  white dwarf models include element diffusion consistently with  evolutionary processes.

The calculations presented here constitute  the first set of
fully  evolutionary  calculations  of  ultra-massive white  dwarfs including realistic initial chemical 
profiles for each white dwarf mass,
an updated microphysics, and the effects of phase separation process duration  crystallization.
All these processes impact to a different extent the cooling times
of ultra-massive white dwarfs. We find a marked dependence of the cooling times with the stellar
mass at low luminosity and a fast cooling in our most massive sequences.
In particular, our $1.29 \, M_\sun$ 
hydrogen-rich sequence reaches $\log(L_\star/L_\sun)= -5$ in 
only 3.6 Gyr, which is even shorter (2.4 Gyr) in the case of the
hydrogen-deficient counterpart. 
Our results also show an enrichment of carbon in the outer layers of the hydrogen-deficient
sequences at 
intermediate luminosities. We have also investigated the effect of element diffusion, 
and found that these processes profoundly change 
the inner abundance distribution from the very early stages of white dwarf evolution.
In particular, the initial helium  and carbon distributions
below the hydrogen-rich envelope result substantially changed when evolution reaches 
low effective temperature, thus impacting the cooling times at 
such advanced stages of evolution.

{ Our new cooling sequences indicate that all  pulsating white dwarfs existing in 
the literature with masses higher than
$1.10 M_\sun$ should have more than 80\% of their mass crystallized if they harbour O-Ne cores.
This is a relevant issue since crystallization has
important consequences on the pulsational properties of massive ZZ Ceti stars. 
This aspect has recently been thoroughly explored in \cite{2018arXiv180703810D}
on the basis of these new sequences,
with relevant implications for the pulsational properties characterizing ultra-massive white dwarfs.}

In summary, we find that  the use  of detailed chemical profiles as given by progenitor evolution
and their time evolution resulting from element 
diffusion processes and from phase separation during crystallization constitute  important
improvements as compared with existing calculations  that 
has to be considered at assessing the cooling times and pulsational properties of
ultra-massive white dwarfs.
We hope that asteroseismological inferences of ultra-massive white dwarfs benefit 
from these new evolutionary sequences, helping to shed light on the crystallization
in the interior of white dwarfs.

\begin{acknowledgements}
This paper is devoted to the memory 
of Enrique Garc\'ia-Berro, which without his
experience, talent and passion, would have not been possible. 
We strongly acknowledge A. Cumming from providing us with
the phase diagram, a key physical ingredient required in our investigation, and  L. Siess for the chemical profiles of 
his models. Part of this work was
supported by AGENCIA through the Programa de Modernizaci\'on Tecnol\'ogica
BID 1728/OC-AR, and by the PIP 112-200801-00940 grant from CONICET. M3B is partially supported through ANPCyT
grant PICT-2016-0053 and MinCyT-DAAD bilateral cooperation program through grant
DA/16/07. 
This research has made use of NASA's Astrophysics Data System. 
\end{acknowledgements}

\bibliographystyle{aa} 
\bibliography{lowZ}

\begin{thebibliography}{64}
\expandafter\ifx\csname natexlab\endcsname\relax\def\natexlab#1{#1}\fi

\bibitem[{{Abrikosov}(1961)}]{Abrikosov1961}
{Abrikosov}, A.~A. 1961, Sov. Phys. JETP, 12, 1254

\bibitem[{{Althaus} {et~al.}(2015){Althaus}, {Camisassa}, {Miller Bertolami},
  {C{\'o}rsico}, \& {Garc{\'{\i}}a-Berro}}]{2015A&A...576A...9A}
{Althaus}, L.~G., {Camisassa}, M.~E., {Miller Bertolami}, M.~M., {C{\'o}rsico},
  A.~H., \& {Garc{\'{\i}}a-Berro}, E. 2015, \aap, 576, A9

\bibitem[{{Althaus} {et~al.}(2010{\natexlab{a}}){Althaus}, {C{\'o}rsico},
  {Bischoff-Kim}, {Romero}, {Renedo}, {Garc{\'{\i}}a-Berro}, \& {Miller
  Bertolami}}]{2010ApJ...717..897A}
{Althaus}, L.~G., {C{\'o}rsico}, A.~H., {Bischoff-Kim}, A., {et~al.}
  2010{\natexlab{a}}, \apj, 717, 897

\bibitem[{{Althaus} {et~al.}(2010{\natexlab{b}}){Althaus}, {C{\'o}rsico},
  {Isern}, \& {Garc{\'{\i}}a-Berro}}]{2010A&ARv..18..471A}
{Althaus}, L.~G., {C{\'o}rsico}, A.~H., {Isern}, J., \& {Garc{\'{\i}}a-Berro},
  E. 2010{\natexlab{b}}, \aapr, 18, 471

\bibitem[{{Althaus} {et~al.}(2012){Althaus}, {Garc{\'{\i}}a-Berro}, {Isern},
  {C{\'o}rsico}, \& {Miller Bertolami}}]{2012A&A...537A..33A}
{Althaus}, L.~G., {Garc{\'{\i}}a-Berro}, E., {Isern}, J., {C{\'o}rsico}, A.~H.,
  \& {Miller Bertolami}, M.~M. 2012, \aap, 537, A33

\bibitem[{{Althaus} {et~al.}(2007){Althaus}, {Garc{\'{\i}}a-Berro}, {Isern},
  {C{\'o}rsico}, \& {Rohrmann}}]{2007A&A...465..249A}
{Althaus}, L.~G., {Garc{\'{\i}}a-Berro}, E., {Isern}, J., {C{\'o}rsico}, A.~H.,
  \& {Rohrmann}, R.~D. 2007, \aap, 465, 249

\bibitem[{{Althaus} {et~al.}(2010{\natexlab{c}}){Althaus},
  {Garc{\'{\i}}a-Berro}, {Renedo}, {Isern}, {C{\'o}rsico}, \&
  {Rohrmann}}]{2010ApJ...719..612A}
{Althaus}, L.~G., {Garc{\'{\i}}a-Berro}, E., {Renedo}, I., {et~al.}
  2010{\natexlab{c}}, \apj, 719, 612

\bibitem[{{Althaus} {et~al.}(2005){Althaus}, {Serenelli}, {Panei},
  {C{\'o}rsico}, {Garc{\'{\i}}a-Berro}, \&
  {Sc{\'o}ccola}}]{2005A&A...435..631A}
{Althaus}, L.~G., {Serenelli}, A.~M., {Panei}, J.~A., {et~al.} 2005, A\&A, 435,
  631

\bibitem[{{Bono} {et~al.}(2013){Bono}, {Salaris}, \&
  {Gilmozzi}}]{2013A&A...549A.102B}
{Bono}, G., {Salaris}, M., \& {Gilmozzi}, R. 2013, \aap, 549, A102

\bibitem[{{Bours} {et~al.}(2015){Bours}, {Marsh}, {G{\"a}nsicke}, {Tauris},
  {Istrate}, {Badenes}, {Dhillon}, {Gal-Yam}, {Hermes}, {Kengkriangkrai},
  {Kilic}, {Koester}, {Mullally}, {Prasert}, {Steeghs}, {Thompson}, \&
  {Thorstensen}}]{2015MNRAS.450.3966B}
{Bours}, M.~C.~P., {Marsh}, T.~R., {G{\"a}nsicke}, B.~T., {et~al.} 2015,
  \mnras, 450, 3966

\bibitem[{{Brassard} \& {Fontaine}(2005)}]{2005ApJ...622..572B}
{Brassard}, P. \& {Fontaine}, G. 2005, \apj, 622, 572

\bibitem[{{Camisassa} {et~al.}(2017){Camisassa}, {Althaus}, {Rohrmann},
  {Garc{\'{\i}}a-Berro}, {Torres}, {C{\'o}rsico}, \&
  {Wachlin}}]{2017ApJ...839...11C}
{Camisassa}, M.~E., {Althaus}, L.~G., {Rohrmann}, R.~D., {et~al.} 2017, \apj,
  839, 11

\bibitem[{{Cassisi} {et~al.}(2007){Cassisi}, {Potekhin}, {Pietrinferni},
  {Catelan}, \& {Salaris}}]{2007ApJ...661.1094C}
{Cassisi}, S., {Potekhin}, A.~Y., {Pietrinferni}, A., {Catelan}, M., \&
  {Salaris}, M. 2007, \apj, 661, 1094

\bibitem[{{Castanheira} {et~al.}(2010){Castanheira}, {Kepler}, {Kleinman},
  {Nitta}, \& {Fraga}}]{2010MNRAS.405.2561C}
{Castanheira}, B.~G., {Kepler}, S.~O., {Kleinman}, S.~J., {Nitta}, A., \&
  {Fraga}, L. 2010, \mnras, 405, 2561

\bibitem[{{Castanheira} {et~al.}(2013){Castanheira}, {Kepler}, {Kleinman},
  {Nitta}, \& {Fraga}}]{2013MNRAS.430...50C}
---. 2013, \mnras, 430, 50

\bibitem[{{C{\'o}rsico} {et~al.}(2005){C{\'o}rsico}, {Althaus}, {Montgomery},
  {Garc{\'{\i}}a-Berro}, \& {Isern}}]{2005A&A...429..277C}
{C{\'o}rsico}, A.~H., {Althaus}, L.~G., {Montgomery}, M.~H.,
  {Garc{\'{\i}}a-Berro}, E., \& {Isern}, J. 2005, \aap, 429, 277

\bibitem[{{C{\'o}rsico} {et~al.}(2004){C{\'o}rsico}, {Garc{\'{\i}}a-Berro},
  {Althaus}, \& {Isern}}]{2004A&A...427..923C}
{C{\'o}rsico}, A.~H., {Garc{\'{\i}}a-Berro}, E., {Althaus}, L.~G., \& {Isern},
  J. 2004, \aap, 427, 923

\bibitem[{{Curd} {et~al.}(2017){Curd}, {Gianninas}, {Bell}, {Kilic}, {Romero},
  {Allende Prieto}, {Winget}, \& {Winget}}]{2017MNRAS.468..239C}
{Curd}, B., {Gianninas}, A., {Bell}, K.~J., {et~al.} 2017, \mnras, 468, 239

\bibitem[{{De Ger{\'o}nimo} {et~al.}(2018){De Ger{\'o}nimo}, {C{\'o}rsico},
  {Althaus}, {Wachlin}, \& {Camisassa}}]{2018arXiv180703810D}
{De Ger{\'o}nimo}, F.~C., {C{\'o}rsico}, A.~H., {Althaus}, L.~G., {Wachlin},
  F.~C., \& {Camisassa}, M.~E. 2018, arXiv e-prints

\bibitem[{{Ferguson} {et~al.}(2005){Ferguson}, {Alexander}, {Allard}, {Barman},
  {Bodnarik}, {Hauschildt}, {Heffner-Wong}, \& {Tamanai}}]{2005ApJ...623..585F}
{Ferguson}, J.~W., {Alexander}, D.~R., {Allard}, F., {et~al.} 2005, \apj, 623,
  585

\bibitem[{{Fontaine} \& {Brassard}(2008)}]{2008PASP..120.1043F}
{Fontaine}, G. \& {Brassard}, P. 2008, \pasp, 120, 1043

\bibitem[{{Garcia-Berro} {et~al.}(1997){Garcia-Berro}, {Isern}, \&
  {Hernanz}}]{1997MNRAS.289..973G}
{Garcia-Berro}, E., {Isern}, J., \& {Hernanz}, M. 1997, \mnras, 289, 973

\bibitem[{{Garc{\'{\i}}a-Berro} {et~al.}(1997){Garc{\'{\i}}a-Berro}, {Ritossa},
  \& {Iben}}]{1997ApJ...485..765G}
{Garc{\'{\i}}a-Berro}, E., {Ritossa}, C., \& {Iben}, Jr., I. 1997, \apj, 485,
  765

\bibitem[{{Garc{\'{\i}}a-Berro} {et~al.}(2010){Garc{\'{\i}}a-Berro}, {Torres},
  {Althaus}, {Renedo}, {Lor{\'e}n-Aguilar}, {C{\'o}rsico}, {Rohrmann},
  {Salaris}, \& {Isern}}]{2010Natur.465..194G}
{Garc{\'{\i}}a-Berro}, E., {Torres}, S., {Althaus}, L.~G., {et~al.} 2010, \nat,
  465, 194

\bibitem[{{Gianninas} {et~al.}(2011){Gianninas}, {Bergeron}, \&
  {Ruiz}}]{2011ApJ...743..138G}
{Gianninas}, A., {Bergeron}, P., \& {Ruiz}, M.~T. 2011, \apj, 743, 138

\bibitem[{{Gil-Pons} {et~al.}(2007){Gil-Pons}, {Guti{\'e}rrez}, \&
  {Garc{\'{\i}}a-Berro}}]{2007A&A...464..667G}
{Gil-Pons}, P., {Guti{\'e}rrez}, J., \& {Garc{\'{\i}}a-Berro}, E. 2007, \aap,
  464, 667

\bibitem[{{Haft} {et~al.}(1994){Haft}, {Raffelt}, \&
  {Weiss}}]{1994ApJ...425..222H}
{Haft}, M., {Raffelt}, G., \& {Weiss}, A. 1994, \apj, 425, 222

\bibitem[{{Hansen} {et~al.}(2013){Hansen}, {Kalirai}, {Anderson}, {Dotter},
  {Richer}, {Rich}, {Shara}, {Fahlman}, {Hurley}, {King}, {Reitzel}, \&
  {Stetson}}]{2013Natur.500...51H}
{Hansen}, B.~M.~S., {Kalirai}, J.~S., {Anderson}, J., {et~al.} 2013, \nat, 500,
  51

\bibitem[{{Hermes} {et~al.}(2013){Hermes}, {Kepler}, {Castanheira},
  {Gianninas}, {Winget}, {Montgomery}, {Brown}, \&
  {Harrold}}]{2013ApJ...771L...2H}
{Hermes}, J.~J., {Kepler}, S.~O., {Castanheira}, B.~G., {et~al.} 2013, \apjl,
  771, L2

\bibitem[{{Iglesias} \& {Rogers}(1996)}]{1996ApJ...464..943I}
{Iglesias}, C.~A. \& {Rogers}, F.~J. 1996, \apj, 464, 943

\bibitem[{{Isern} {et~al.}(1997){Isern}, {Mochkovitch}, {Garc{\'{\i}}a-Berro},
  \& {Hernanz}}]{1997ApJ...485..308I}
{Isern}, J., {Mochkovitch}, R., {Garc{\'{\i}}a-Berro}, E., \& {Hernanz}, M.
  1997, \apj, 485, 308

\bibitem[{{Itoh} {et~al.}(1994){Itoh}, {Hayashi}, \&
  {Kohyama}}]{1994ApJ...436..418I}
{Itoh}, N., {Hayashi}, H., \& {Kohyama}, Y. 1994, \apj, 436, 418

\bibitem[{{Itoh} {et~al.}(1996){Itoh}, {Hayashi}, {Nishikawa}, \&
  {Kohyama}}]{1996ApJS..102..411I}
{Itoh}, N., {Hayashi}, H., {Nishikawa}, A., \& {Kohyama}, Y. 1996, \apjs, 102,
  411

\bibitem[{{Jeffery} {et~al.}(2011){Jeffery}, {von Hippel}, {DeGennaro}, {van
  Dyk}, {Stein}, \& {Jefferys}}]{2011ApJ...730...35J}
{Jeffery}, E.~J., {von Hippel}, T., {DeGennaro}, S., {et~al.} 2011, \apj, 730,
  35

\bibitem[{{Jim{\'e}nez-Esteban} {et~al.}(2018){Jim{\'e}nez-Esteban}, {Torres},
  {Rebassa-Mansergas}, {Skorobogatov}, {Solano}, {Cantero}, \&
  {Rodrigo}}]{2018MNRAS.480.4505J}
{Jim{\'e}nez-Esteban}, F.~M., {Torres}, S., {Rebassa-Mansergas}, A., {et~al.}
  2018, \mnras, 480, 4505

\bibitem[{{Kanaan} {et~al.}(1992){Kanaan}, {Kepler}, {Giovannini}, \&
  {Diaz}}]{1992ApJ...390L..89K}
{Kanaan}, A., {Kepler}, S.~O., {Giovannini}, O., \& {Diaz}, M. 1992, \apjl,
  390, L89

\bibitem[{{Kanaan} {et~al.}(2005){Kanaan}, {Nitta}, {Winget}, {Kepler},
  {Montgomery}, {Metcalfe}, {Oliveira}, {Fraga}, {da Costa}, {Costa},
  {Castanheira}, {Giovannini}, {Nather}, {Mukadam}, {Kawaler}, {O'Brien},
  {Reed}, {Kleinman}, {Provencal}, {Watson}, {Kilkenny}, {Sullivan},
  {Sullivan}, {Shobbrook}, {Jiang}, {Ashoka}, {Seetha}, {Leibowitz},
  {Ibbetson}, {Mendelson}, {Mei{\v s}tas}, {Kalytis}, {Ali{\v s}auskas},
  {O'Donoghue}, {Buckley}, {Martinez}, {van Wyk}, {Stobie}, {Marang}, {van
  Zyl}, {Ogloza}, {Krzesinski}, {Zola}, {Moskalik}, {Breger}, {Stankov},
  {Silvotti}, {Piccioni}, {Vauclair}, {Dolez}, {Chevreton}, {Deetjen},
  {Dreizler}, {Schuh}, {Gonzalez Perez}, {{\O}stensen}, {Ulla}, {Manteiga},
  {Suarez}, {Burleigh}, \& {Barstow}}]{2005A&A...432..219K}
{Kanaan}, A., {Nitta}, A., {Winget}, D.~E., {et~al.} 2005, \aap, 432, 219

\bibitem[{{Kepler} {et~al.}(2016){Kepler}, {Pelisoli}, {Koester}, {Ourique},
  {Romero}, {Reindl}, {Kleinman}, {Eisenstein}, {Valois}, \&
  {Amaral}}]{2016MNRAS.455.3413K}
{Kepler}, S.~O., {Pelisoli}, I., {Koester}, D., {et~al.} 2016, \mnras, 455,
  3413

\bibitem[{{Kirzhnits}(1960)}]{Kirzhnits1960}
{Kirzhnits}, D.~A. 1960, Sov. Phys. JETP, 11, 365

\bibitem[{{Kleinman} {et~al.}(2013){Kleinman}, {Kepler}, {Koester}, {Pelisoli},
  {Pe{\c c}anha}, {Nitta}, {Costa}, {Krzesinski}, {Dufour}, {Lachapelle},
  {Bergeron}, {Yip}, {Harris}, {Eisenstein}, {Althaus}, \&
  {C{\'o}rsico}}]{2013ApJS..204....5K}
{Kleinman}, S.~J., {Kepler}, S.~O., {Koester}, D., {et~al.} 2013, \apjs, 204, 5

\bibitem[{{Magni} \& {Mazzitelli}(1979)}]{1979A&A....72..134M}
{Magni}, G. \& {Mazzitelli}, I. 1979, \aap, 72, 134

\bibitem[{{Maoz} {et~al.}(2014){Maoz}, {Mannucci}, \&
  {Nelemans}}]{2014ARA&A..52..107M}
{Maoz}, D., {Mannucci}, F., \& {Nelemans}, G. 2014, \araa, 52, 107

\bibitem[{{Medin} \& {Cumming}(2010)}]{2010PhRvE..81c6107M}
{Medin}, Z. \& {Cumming}, A. 2010, \pre, 81, 036107

\bibitem[{{Metcalfe} {et~al.}(2004){Metcalfe}, {Montgomery}, \&
  {Kanaan}}]{2004ApJ...605L.133M}
{Metcalfe}, T.~S., {Montgomery}, M.~H., \& {Kanaan}, A. 2004, \apjl, 605, L133

\bibitem[{{Miller Bertolami}(2016)}]{2016A&A...588A..25M}
{Miller Bertolami}, M.~M. 2016, \aap, 588, A25

\bibitem[{{Montgomery} \& {Winget}(1999)}]{1999ApJ...526..976M}
{Montgomery}, M.~H. \& {Winget}, D.~E. 1999, \apj, 526, 976

\bibitem[{{Mukadam} {et~al.}(2004){Mukadam}, {Mullally}, {Nather}, {Winget},
  {von Hippel}, {Kleinman}, {Nitta}, {Krzesi{\'n}ski}, {Kepler}, {Kanaan},
  {Koester}, {Sullivan}, {Homeier}, {Thompson}, {Reaves}, {Cotter},
  {Slaughter}, \& {Brinkmann}}]{2004ApJ...607..982M}
{Mukadam}, A.~S., {Mullally}, F., {Nather}, R.~E., {et~al.} 2004, \apj, 607,
  982

\bibitem[{{Nitta} {et~al.}(2016){Nitta}, {Kepler}, {Chen{\'e}}, {Koester},
  {Provencal}, {Kleinmani}, {Sullivan}, {Chote}, {Sefako}, {Kanaan}, {Romero},
  {Corti}, {Kilic}, {Montgomery}, \& {Winget}}]{2016IAUFM..29B.493N}
{Nitta}, A., {Kepler}, S.~O., {Chen{\'e}}, A.-N., {et~al.} 2016, IAU Focus
  Meeting, 29, 493

\bibitem[{{Rebassa-Mansergas} {et~al.}(2015){Rebassa-Mansergas}, {Rybicka},
  {Liu}, {Han}, \& {Garc{\'{\i}}a-Berro}}]{2015MNRAS.452.1637R}
{Rebassa-Mansergas}, A., {Rybicka}, M., {Liu}, X.-W., {Han}, Z., \&
  {Garc{\'{\i}}a-Berro}, E. 2015, \mnras, 452, 1637

\bibitem[{{Renedo} {et~al.}(2010){Renedo}, {Althaus}, {Miller Bertolami},
  {Romero}, {C{\'o}rsico}, {Rohrmann}, \&
  {Garc{\'{\i}}a-Berro}}]{2010ApJ...717..183R}
{Renedo}, I., {Althaus}, L.~G., {Miller Bertolami}, M.~M., {et~al.} 2010, \apj,
  717, 183

\bibitem[{{Rohrmann}(2018)}]{2018MNRAS.473..457R}
{Rohrmann}, R.~D. 2018, \mnras, 473, 457

\bibitem[{{Rohrmann} {et~al.}(2012){Rohrmann}, {Althaus},
  {Garc{\'{\i}}a-Berro}, {C{\'o}rsico}, \& {Miller
  Bertolami}}]{2012A&A...546A.119R}
{Rohrmann}, R.~D., {Althaus}, L.~G., {Garc{\'{\i}}a-Berro}, E., {C{\'o}rsico},
  A.~H., \& {Miller Bertolami}, M.~M. 2012, \aap, 546, A119

\bibitem[{{Salaris} {et~al.}(2013){Salaris}, {Althaus}, \&
  {Garc{\'{\i}}a-Berro}}]{2013A&A...555A..96S}
{Salaris}, M., {Althaus}, L.~G., \& {Garc{\'{\i}}a-Berro}, E. 2013, \aap, 555,
  A96

\bibitem[{{Salaris} {et~al.}(2009){Salaris}, {Serenelli}, {Weiss}, \& {Miller
  Bertolami}}]{2009ApJ...692.1013S}
{Salaris}, M., {Serenelli}, A., {Weiss}, A., \& {Miller Bertolami}, M. 2009,
  \apj, 692, 1013

\bibitem[{{Segretain} {et~al.}(1994){Segretain}, {Chabrier}, {Hernanz},
  {Garc\'ia-Berro}, {Isern}, \& {Mochkovitch}}]{1994ApJ...434..641S}
{Segretain}, L., {Chabrier}, G., {Hernanz}, M., {et~al.} 1994, \apj, 434, 641

\bibitem[{{Siess}(2006)}]{2006A&A...448..717S}
{Siess}, L. 2006, \aap, 448, 717

\bibitem[{{Siess}(2007)}]{2007A&A...476..893S}
---. 2007, \aap, 476, 893

\bibitem[{{Siess}(2010)}]{2010A&A...512A..10S}
---. 2010, \aap, 512, A10

\bibitem[{{Tassoul} {et~al.}(1990){Tassoul}, {Fontaine}, \&
  {Winget}}]{1990ApJS...72..335T}
{Tassoul}, M., {Fontaine}, G., \& {Winget}, D.~E. 1990, \apjs, 72, 335

\bibitem[{{Tremblay} {et~al.}(2019){Tremblay}, {Cukanovaite}, {Gentile
  Fusillo}, {Cunningham}, \& {Hollands}}]{2019MNRAS.482.5222T}
{Tremblay}, P.-E., {Cukanovaite}, E., {Gentile Fusillo}, N.~P., {Cunningham},
  T., \& {Hollands}, M.~A. 2019, \mnras, 482, 5222

\bibitem[{{van Horn}(1968)}]{1968ApJ...151..227V}
{van Horn}, H.~M. 1968, \apj, 151, 227

\bibitem[{{Weiss} \& {Ferguson}(2009)}]{2009A&A...508.1343W}
{Weiss}, A. \& {Ferguson}, J.~W. 2009, \aap, 508, 1343

\bibitem[{{Winget} \& {Kepler}(2008)}]{2008ARA&A..46..157W}
{Winget}, D.~E. \& {Kepler}, S.~O. 2008, \araa, 46, 157

\bibitem[{{Winget} {et~al.}(2009){Winget}, {Kepler}, {Campos}, {Montgomery},
  {Girardi}, {Bergeron}, \& {Williams}}]{2009ApJ...693L...6W}
{Winget}, D.~E., {Kepler}, S.~O., {Campos}, F., {et~al.} 2009, \apjl, 693, L6

\end{thebibliography}
 
\end{document}